\documentclass{emulateapj}

\usepackage{graphicx}
\usepackage{amssymb}
\usepackage{longtable}
\usepackage{epsf}
\usepackage{apjfonts}
\usepackage{rotating}
\usepackage{float}
\usepackage{adjustbox}

\usepackage{natbib}
\slugcomment{Submitted to the Astrophysical Journal}

\shorttitle{Spectroscopy of ultra massive galaxies}
\shortauthors{Marsan et al.}

\begin{document}

\title{A spectroscopic follow-up program of very massive galaxies at 3<z<4: confirmation of spectroscopic redshifts, and a high fraction of powerful AGN}

\author{Z.~Cemile~Marsan\altaffilmark{1}, Danilo~Marchesini\altaffilmark{1}, 
Gabriel B. Brammer\altaffilmark{2}, 
Stefan Geier\altaffilmark{3},
Erin Kado-Fong\altaffilmark{1},
Ivo Labb\'e\altaffilmark{4}, 
Adam Muzzin\altaffilmark{5},
Mauro Stefanon\altaffilmark{4}}

\altaffiltext{1}{Department of Physics and Astronomy, Tufts University, 
574 Boston Avenue, Medford, MA 02155, USA}
\altaffiltext{2}{Space Telescope Science Institute, 3700 San Martin Drive, 
  Baltimore, MD 21218, USA}
\altaffiltext{3}{Instituto de Astrof\'{\i}sica de Canarias (IAC), E-38205 La Laguna, Tenerife, Spain}
\altaffiltext{4}{Leiden Observatory, Leiden University, PO Box 9513, NL-2300 
RA Leiden, The Netherlands}
\altaffiltext{5}{Department of Physics and Astronomy, York University, 
 4700 Keele St., Toronto, Ontario, MJ3 1P3, Canada}

\begin{abstract}
We present the analysis and results of a spectroscopic follow-up program of a mass-selected sample of six galaxies at $3<z<4$ using data from Keck-NIRPSEC and VLT-Xshooter. We confirm the $z>3$ redshifts for half of the sample through the detection of strong nebular emission lines, and improve the $z_{phot}$ accuracy for the remainder of the sample through the combination of photometry and spectra. The modeling of the emission-line-corrected spectral energy distributions (SEDs) adopting improved redshifts confirms the very large stellar masses of the sample ($M_{*}\sim 1.5 - 4 \times 10^{11} M_{\odot}$) in the first 2 Gyrs of cosmic history, with a diverse range in stellar ages, star formation rates and dust content. From the analysis of emission line luminosities and widths, and far-infrared (FIR) fluxes we confirm that $\gtrsim 80\%$ of the sample are hosts to luminous hidden active galactic nuclei (AGNs), with bolometric luminosities of $\sim 10^{44-46}$ erg~s$^{-1}$. We find that the MIPS~$24\mu$m photometry is largely contaminated by AGN continuum, rendering the SFRs derived using only $24\mu$m photometry to be severely overestimated. By including the emission from the AGN in the modeling of the UV-to-FIR SEDs, we confirm that the presence of the AGN does not bias considerably the stellar masses ($<0.3$ dex at 1$\sigma$). We show evidence for a rapid increase of the AGN fraction from $\sim$30\% to $\sim$60-100\% over the 1~Gyr between $z\sim2$ and $z\sim3$. Although we cannot exclude some enhancement of the AGN fraction for our sample due to selection effects, the small measured [OIII] contamination to the observed K-band fluxes suggests that our sample is not significantly biased toward massive galaxies hosting AGNs.

\end{abstract}



\section{INTRODUCTION}\label{sec-in}

One of the surprising findings in observational studies of 
galaxies in the early universe
is that, in contrast to the bottom-up assembly of dark matter halos inferred 
from simulations of $\Lambda$CDM cosmologies, in which low-mass halos form 
early and subsequently grow via continued accretion and merging to form more 
massive halos at later times, the stellar component of halos appear to have assembled in an anti-hierarchical, top-down manner, with low-mass galaxies assembling most of their mass and forming most of their stars at later times compared to more massive systems \citep{thomas05,gallazzi06,perezgonzalez08, wiklind08, marchesini09, marchesini10, dominguezsanchez11, muzzin13b,stefanon15}. 

Closely related to this issue is the intriguing finding that the 
number density of the most massive galaxies (i.e., those with stellar mass M$_{*} > 3\times 10^{11}M_{\odot}$) seems to 
evolve slowly from $z\sim$4 to $z\sim$1.5 
\citep{perezgonzalez08, marchesini09, marchesini10, muzzin13b}, suggesting that 
very massive galaxies were already in place at z $\sim$ 3.5, and 
implying that their stellar content was assembled rapidly in 
the first $\sim$1.5 Gyr of cosmic history \citep{stefanon15, caputi15}. 
 While theoretical models of galaxy formation
 and evolution have 
 significantly improved in matching observations in recent years, they still are unable to reproduce
 the observed number density of galaxies at 
 3 < z < 4 with log (M$_{*}$/M$_{\odot}$) > 11.5
 (\citealt{fontanot09, marchesini09, marchesini10, ilbert13}), unless the model-predicted stellar mass functions (SMFs) are convolved to account for the effect of Eddington bias by $\sim$0.3-0.4 dex \citep{henriques15}, arguably much larger than the typical random uncertainties on $M_{*}$ at these redshifts (e.g., \citealt{marchesini10, ilbert13, muzzin13b}). Furthermore, it remains to be determined to what extent the inferred 
number density measurements at high $z$ are affected by blending of sources 
in ground-based images.

In recent years, robust evidence for the existence of massive (M$_{*} \gtrsim 10^{11}$ M$_{\odot}$) galaxies at $z>$3 was enabled by the well-sampled spectral energy distributions (SEDs) and accurate photometric redshifts delivered by near-infrared (NIR) imaging surveys adopting medium bandwidth filters in the NIR, i.e., NEWFIRM Medium-Band Survey (NMBS; \citealt{whitaker11}, and the FourStar Galaxy Evolution Survey (zFOURGE; \citealt{spitler14}). The wide-field NMBS robustly revealed, a population of `monster' galaxies at $3<z<4$ with log($M_{*}/M_{\odot}$)>11.4 \citep{marchesini10}, whereas the deeper but pencil beam zFOURGE survey extended the study of massive galaxies at $z>3$ down to the characteristic stellar mass at $z\sim$3-4, i.e., log($M_{*}/M_{\odot}$)$\sim$10.6 \citep{spitler14, straatman14}. \citet{stefanon15} used the UltraVISTA DR1 \citep{muzzin13a} to search for very massive galaxies at $4<z<7$, finding a robust candidate of a monster galaxy at $z\sim$5.6 with log($M_{*}/M_{\odot}$)$\sim$11.6 and quenched star-formation activity.

In contrast to the local universe where massive galaxies are predominantly quiescent and have 
old stellar populations with red colors \citep{blanton03}, the massive galaxy 
population at $z\sim3$ is dominated by dusty, star-forming galaxies ($\sim 40-60\%$ of massive population, \citealt{marchesini10, spitler14}), and includes a significant population of already quiescent galaxies ($\sim30-40\%$) with ages consistent with the age of the universe at the targeted redshifts. In addition, from the X-ray and radio detections, as well as the ubiquitous very bright fluxes in MIPS 24$\mu$m (rest-frame 5-6 $\mu$m), it appears that massive galaxies in the early universe commonly host active galactic nuclei, both at $z>3$ \citep{marchesini10, stefanon15}, as well as at $z<3$ \citep{barro13, forsterschreiber14}.

Despite the well-sampled SEDs delivered by the NMBS and zFOURGE surveys, there is still ambiguity in the photometric redshift solutions for the rest-frame UV faint massive galaxies at $z>3$. Quantitatively, among the population of very massive galaxies with log($M_{*}/M_{\odot}$)>11.4 at $3<z<4$ selected from the NMBS by \citet{marchesini10}, $\sim50\%$ could have a photometric redshift $z_{\rm phot}<3$ solution if their stellar populations were characterized by both an evolved (i.e, stellar age $\gtrsim$ 1 Gyr) and very dusty (i.e., $<A_{\rm V}> \sim 3$ mag) component. This elusive population of high-$z$ old and dusty galaxies, originally proposed in \citet{marchesini10} and effectively non existent at $z<1$, appears to play a major role among high-z massive galaxies \citep{marchesini14}, and has only recently been targeted for detailed investigations (e.g., \citealt{bedregal15, brammer17}). 

Because of the ambiguity in the photometric redshift solutions for as much as half of the population, it is of paramount importance to measure spectroscopically the redshifts of very massive galaxies at $z>3$. Recently, \citet{marsan15} presented the first spectroscopic confirmation of an ultra-massive (M$_{*}\sim3\times10^{11}$ M$_{\odot}$) galaxy at $z=3.35$, dubbed 'The Vega Galaxy', due to the incredible similarity of the integrated SED to an A0V-star such as Vega. The detailed analysis of its UV-to-FIR SED reveals that most of its stars formed at $z>4$ in a highly dissipative, intense, and short burst of star formation, and that it is transitioning to a post-starburst phase while hosting a powerful AGN.

In this paper, we present the results of a spectroscopic follow-up program
of a stellar-mass-complete sample of 6 galaxies at $3 <z< 4$ originally selected 
from the NMBS \citep{marchesini10}, aimed at
spectroscopically confirming their redshifts and further investigating 
their stellar populations. The brightest galaxy within this sample has been 
extensively studied and the results were presented in \citet{marsan15}. 

We confirm $z>3$ redshifts for half of the sample through the detection of nebular emission lines, 
and improve the $z_{phot}$ accuracy for the remainder of the sample with the combination of photometry 
and spectra. We present
the SED modeling of the combined NMBS photometry
and spectroscopy and discuss the properties of the mass complete sample at $z>3$. 

This paper is organized as follows: In \S\ref{sec-ss} we describe the target selection, and present the ground-based spectroscopy and data reduction of the mass selected sample in \S\ref{sec-data}. 
In \S\ref{sec-analysis}, we present the analysis of the 
spectra, update the redshifts for our targets ($z_{spec}$, if emission lines are detected; otherwise $z_{cont}$, obtained through including binned spectra in the modeling of the observed spectral energy distributions, SEDs), the modeling of the SEDs with updated redshifts, and the investigation of the active galactic nuclei (AGN) content of the sample. The results are summarized and discussed 
in \S\ref{sec-disc}. We assume $\Omega_{\rm M}=0.3$, $\Omega_{\rm \Lambda}=0.7$, 
$H_{\rm 0}=70$~km~s$^{-1}$~Mpc$^{-1}$, and a \citet{kroupa01} initial mass 
function (IMF) throughout the paper. All magnitudes are in the AB system. 

\section{Target Selection}\label{sec-ss}
The targets presented in this paper are primarily selected from the stellar mass complete 
sample of massive galaxies log(M$_{*}/$M$_{\odot}$)>11.4 at 3 < $z$ < 4 presented in 
\citet{marchesini10}, constructed using the NMBS \citep{whitaker11}. The 
NMBS is a moderately wide, moderately deep near-infrared imaging survey 
\citep{vandokkum09} targeting the COSMOS field \citep{scoville07} and 
AEGIS strip \citep{davis07}. The NMBS photometry presented in 
\citet{whitaker11} includes deep optical 
\emph{ugriz} data from the CFHT Legacy Survey, deep 
\emph{Spitzer}-IRAC and MIPS imaging \citep{sanders07}, \emph{Galaxy Evolution 
Explorer} ({\it GALEX}) photometry in the FUV (150 nm) and NUV (225 nm) 
passbands \citep{martin05}, NIR imaging with NEWFIRM using the five 
medium-band filters $J_{\rm 1}$, $J_{\rm 2}$, $J_{\rm 3}$, $H_{\rm 1}$, 
$H_{\rm 2}$, broad-band $K$ \citep{vandokkum09} for both fields. 
The COSMOS field additionally includes deep Subaru images 
with the $B_{J}V_{J}r^{+}i^{+}z^{+}$ broadband filters \citep{capak07}, Subaru 
images with 12 optical intermediate-band filters from 427 to 827~nm 
\citep{taniguchi07},  and $JHK_{\rm S}$ broad-band imaging from the WIRCam Deep 
Survey (WIRDS; \citealt{mccracken10}). 

The medium-bandwidth NIR filters in NMBS allows the Balmer/4000\AA ~break of
galaxies at 1.5 < $z$ < 3.5 to be more finely sampled compared to the standard
broadband NIR filters 
providing more accurate photometric redshift estimates \citep{whitaker11}. 
The public NMBS catalog uses photometric redshifts determined with the EAZY code \citep{brammer08}
modeling the full FUV-8$\mu$m SEDs. Stellar mass and other
stellar population parameters were determined using Fitting and Assessment of Synthetic
Templates (FAST; \citealt{kriek09}), adopting the redshift solution output by EAZY (i.e., $z_{peak}$)
, the stellar
population synthesis models of \citet{bc03}, the \citet{calzetti00} reddening law
with $A_{\rm V}$ = 0-4 mag, solar metallicity, and
an exponentially declining star formation history (SFH). These SED-modeling assumptions are the same used in the analysis presented in \citet{marchesini10}

Table~\ref{tab-obs} lists the properties of the 6 targets selected for spectroscopic follow-up. Four objects are the brightest candidates of very massive galaxies at $z>3$ selected from \citet{marchesini10}, namely C1-15182, C1-19536, A2-15753, and C1-23152 (i.e., the Vega galaxy, already studied in detail in \citealt{marsan15}). The spectroscopic program targeted two additional galaxies, C1-2127 and C1-19764, consistent with being at $z\gtrsim$3 but not originally included in the sample presented in \citet{marchesini10} because they did not strictly satisfy the completeness limit in stellar mass (though both massive with log(M$_{*}$/M$_{\odot}$)$\sim$11.2-11.3). Table~\ref{tab-obs} also lists C1-21316 from the sample of \citet{marchesini10} but not targeted by our spectroscopic follow-up program, since it has a previously confirmed spectroscopic redshift of $z_{spec}$=3.971 (\citealt{capak10, smolcic12}; but see also \citealt{miettinen15}). C1-21316 is also a sub-millimeter galaxy (SMG) listed as AzTEC 5 in the AzTEC millimeter survey catalog in COSMOS \citep{scott08}. 


\section{Observations and Data Reduction}\label{sec-data}
In this section we briefly describe the ground-based spectroscopic data  
obtained from Keck-NIRSPEC and VLT-X-shooter as part of follow-up 
programs aimed at spectroscopically confirming the existence of very massive 
galaxies at $z>3$. Along with the observational techniques, we provide a 
summary of the data reduction. We refer the reader to 
\citet{geier13} and \citet{marsan15} for a detailed description of the 
spectroscopic data reduction. 
Table~\ref{tab-obs} summarizes the wavelength coverage, total 
exposure time, and seeing of the 
spectroscopic observations. We note that several targets were observed 
with both NIRSPEC and X-Shooter (C1-15182, C1-19536, C1-19764 and C1-23152).

\begin{deluxetable*}{lccccccccccc}
\centering
\tablecaption{Ground-based Spectroscopic Observations\label{tab-obs}}
\tablehead{  & & &   & \colhead{NMBS} &  & &\colhead{NIRSPEC}& & & \colhead{X-shooter} & \\ 
\colhead{ID} & \colhead{$\alpha$}(J2000) & \colhead{$\delta$}(J2000) & \colhead{$K$} & 
\colhead{$z_{peak}$} & \colhead{log $M_{*}$}& \colhead{$\lambda_{\rm range}$} & 
\colhead{$t_{\rm exp}$} & \colhead{FWHM} & \colhead{$\lambda_{\rm range}$} & \colhead{$t_{\rm exp}$} &
\colhead{FWHM}\\
 &  &  & (mag) & & ($M_{\odot}$)& ($\mu$) &  ($min$) & ($^{\prime \prime}$) &  ($\mu$) &  ($min$) & ($^{\prime \prime}$) } 
\startdata
C1-2127  & $10^{\rm h}00^{\rm m}12^{\rm s}.96$& $+02^{\rm d}12^{\rm m}11^{\rm s}.5$  
& 21.69$\pm$0.07 & 3.17$^{+0.12}_{-0.12}$ & 11.26$^{+0.13}_{-0.15}$ & 1.94 - 2.37 & 60  & 0.5 & --      & -- & -- \\
C1-15182 & $09^{\rm h}59^{\rm m}24^{\rm s}.39$ & $+02^{\rm d}25^{\rm m}36^{\rm s}.5$ 
& 21.62$\pm$0.09 & 3.56$^{+0.11}_{-0.11}$ & 11.54$^{+0.04}_{-0.05}$ & 2.04 - 2.46 & 60  & 0.5 & 0.3-2.4 & 60 & 0.5 \\
C1-19536 & $09^{\rm h}59^{\rm m}31^{\rm s}.82$ & $+02^{\rm d}30^{\rm m}18^{\rm s}.2$  
& 21.65$\pm$0.06 & 3.19$^{+0.07}_{-0.08}$ & 11.55$^{+0.03}_{-0.03}$ & 1.94 - 2.37 & 60  & 0.5 & 0.3-2.4 & 60 & 0.8 \\
C1-19764 & $10^{\rm h}00^{\rm m}21^{\rm s}.12$& $+02^{\rm d}30^{\rm m}33^{\rm s}.5$ 
& 21.81$\pm$0.12 & 3.04$^{+0.17}_{-0.19}$ & 11.22$^{+0.03}_{-0.06}$ & 1.94 - 2.37 & 60  & 0.5 & 0.3-2.4 & 60 & 1.1 \\
C1-21316$^a$ & $10^{\rm h}00^{\rm m}19^{\rm s}.74$& $+02^{\rm d}32^{\rm m}04^{\rm s}.3$  
& 22.29$\pm$0.16 & 3.68$^{+0.12}_{-0.11}$ & 11.52$^{+0.01}_{-0.61}$ & -- & --  & --  & --      & -- & -- \\
C1-23152$^b$ & $10^{\rm h}00^{\rm m}27^{\rm s}.81$& $+02^{\rm d}33^{\rm m}49^{\rm s}.3$
& 20.31$\pm$0.02 & 3.29$^{+0.06}_{-0.06}$ & 11.42$^{+0.01}_{-0.01}$ & 1.48 - 2.37 & 75  & 0.7 & 0.3-2.4 & 60 & 0.5 \\
A2-15753 & $14^{\rm h}18^{\rm m}30^{\rm s}.83$&$ +52^{\rm d}40^{\rm m}24^{\rm s}.6$ 
& 22.25$\pm$0.06 & 3.14$^{+0.10}_{-0.09}$ & 11.40$^{+0.02}_{-0.09}$ & 1.94 - 2.37 & 210 & 0.6 & --      & -- & -- \\

\enddata
\tablecomments{Listed ID's are from the NMBS catalog (v4.4). 'C1-' and 'A2-'
refer to the COSMOS and AEGIS fields, respectively. The listed redshifts are the best-fit EAZY redshifts ($z_{peak}$), quoted stellar masses are the best-fit FAST stellar masses used for the selection of the targets (from the NMBS catalog v4.4, \citealt{marchesini10}). $\lambda_{\rm range}$ is the wavelength range covered by the 
instrumental setup; $t_{\rm exp}$ is the on-source exposure time in minutes; 
FWHM is the average seeing in arcsec of the observations. $^a$ C1-21316 was not included in the spectroscopic follow-up program, but it has a spectroscopic redshift present in \citet{capak10}. $^b$ A detailed study of C1-23152 was already presented in \citet{marsan15}. } 
\end{deluxetable*}


\subsection{NIRSPEC Spectroscopy}

We used NIRSPEC \citep{mclean98} on the Keck II telescope for H and K band spectroscopy 
of massive $z>3$ targets primarily in search of their [OIII] emission lines. The 
observations were carried out on the nights of February 11-13, 2011 as part of the 
NOAO program 2011A-0514 (PI: Marchesini) with a typical seeing of 
0.7$^{\prime \prime}$ which worsened throughout the sequence of observations 
due to cloudy variable weather. Observations were conducted following an 
ABA$^{\prime}$B$^{\prime}$ on-source dither pattern. The orientation of the slit 
was set to include a bright point source when possible 
to serve as reference when analyzing 
and combining the two dimensional rectified frames. Targets were acquired 
using blind offsets from a nearby bright star. The alignment of the offset 
star in the slit was checked before each individual 900 sec science exposure and 
corrected when necessary. Before and after each observing sequence, a 
spectrophotometric standard and an AV0 star was observed for the purpose of 
correcting for telluric absorption and detector response. 

The data reduction for the NIRSPEC observations used a combination of custom 
IDL scripts  and standard IRAF tasks\footnote{IRAF is distributed by the 
National Optical Astronomy Observatory, which is operated by the
Association of Universities for Research in Astronomy (AURA), Inc., 
under cooperative agreement with the National Science Foundation.}. 
Bad pixel masks 
were created by flagging outlier pixels in dark and flat frames. The cosmic 
rays on the science frames were removed using L.A.Cosmic \citep{vandokkum01}. 
Each spectrum was sky subtracted using 
an adjacent spectrum with the IDL routines written by George Becker (private communication). The sky 
subtracted frames and the sky spectra were rotated and rectified to a linear wavelength 
scale using a polynomial to interpolate between adjacent pixels.
An absolute wavelength dispersion solution was applied with the use of
OH skylines. 
The standard star frames used to correct for atmospheric absorption and 
detector response were rectified and reduced in the same manner as the science 
frames. 
A one-dimensional spectrum was extracted for each telluric standard 
star (before and after science observations) by summing all the rows (along 
the spatial direction) with a flux greater than 0.1 times that of the central 
row. The average of the one-dimensional telluric star spectra was used to 
correct the two dimensional science and spectrophotometric star frames for 
telluric absorption. The one-dimensional spectrum of the spectrophotometric 
star was extracted in the same manner, and used to create a response function 
for flux calibration. The position of an on-slit bright source was used to 
determine the necessary shifts to align target continuum. 
The two dimensional rectified and reduced science frames 
were combined by weighting according to their signal-to-noise ratio (S/N).

\subsubsection{VLT-X-shooter}
X-shooter is a single-object, medium-resolution echelle spectrograph with 
simultaneous coverage of wavelength range 0.3-2.5$\mu$m in three arms (UVB, 
VIS, NIR) \citep{dodorico06}. The observations of our targets were carried out 
in queue mode as part of the ESO program 087.A-0514 (PI: Brammer) in May 2011, following an 
ABA$^{\prime}$B$^{\prime}$ on-source dither pattern using the 
11$^{\prime \prime}$ $\times$1.0$^{\prime \prime}$ and 
11$^{\prime \prime}$ $\times$0.9$^{\prime \prime}$ slits for the UVB and VIS/NIR 
arms, respectively. This instrumental setup resulted in a spectral resolution 
of R=4200, 8250, and 4000 for the NIR, VIS and UVB arms, respectively. For calibration
purposes, telluric and spectrophotometric standard stars were observed in the same setup 
as science observations. We refer the reader to Table~\ref{tab-obs} for the seeing conditions (FHWM, $^{\prime\prime}$) and exposure times for each object. 
The data reduction for the X-shooter observations used custom scripts based on 
the standard X-shooter reduction pipeline \citep{modigliani10}. The calibration 
steps (master darks, order prediction, flat fields, and the two-dimensional 
maps for later rectification of the spectra) were run with the default 
parameters in the pipeline \citep{goldoni11}. After these five calibration 
steps, the echelle spectra were dark-subtracted, flat-fielded, and rectified, 
and the orders stitched (12, 15 and 16 for the UVB, VIS and NIR arms, 
respectively). The sky was then subtracted using adjacent exposures. Standard 
star observations were reduced with the same calibration data as the science 
frames and used to correct for  telluric absorption and  detector response. 
A final spectrum was created by mean stacking all exposures. We refer to 
\citet{geier13} for a detailed description of reduction steps of X-shooter 
spectra.

\subsubsection{Extraction of One-dimensional Spectra}
Following \citet{horne86}, one-dimensional spectra were extracted by summing 
all adjacent lines (along the spatial direction) using weights corresponding 
to a Gaussian centered on the central row with a full width at half maximum 
equal to the slit width used in each observation. To correct for slit 
losses and obtain an absolute flux calibration, spectroscopic broad/medium-band 
fluxes were obtained by integrating over the corresponding filter curves, and 
a constant scaling was applied to each spectra individually. A binned, lower 
resolution spectrum with higher S/N was extracted for each spectra using 
optimal weighting, excluding parts of spectra contaminated by strong sky 
emission or strong atmospheric absorption. The resulting spectral resolutions 
of the binned X-shooter spectra were $R\approx 10-45$, $15-40$, and $25-50$ for 
the UVB, VIS, and NIR arms, respectively, while the spectral resolutions of 
the binned NIRSPEC spectra were $R\approx 30-100$ and $30-300$ for the $H$ 
and $K$ bands, respectively.


\section{ANALYSIS and Results} \label{sec-analysis}

In this section we built and improved upon previously 
derived and published stellar population parameters
from \citet{marchesini10} and \citet{whitaker11} by modeling the emission-line corrected SEDs constructed from the combination of the UV-to-8 $\mu$m NMBS photometry and the binned spectra. We adopted the measured spectroscopic redshift, $z_{spec}$, when available, or $z_{cont}$, the improved photometric redshift estimated by modeling the binned spectrum in combination with the UV-to-8 $\mu$m NMBS photometry. [OIII] nebular emission lines were detected in the two brightest targets, C1-19536 and C1-15182, allowing for the measurement of the spectroscopic redshift. For the remaining targets with
continuum detections (C1-2127, A2-15753, and C1-19764), we combined the binned spectrum and the NMBS photometry to model the finely sampled SEDs with EAZY to derive $z_{cont}$. In addition to the EAZY template set adopted in \citet{whitaker11}, we also modeled the SEDs using a template set augmented by an 'old-and-dusty' template consisting of a 1 Gyr old single stellar population with $A_{\rm V}$=3 mag of dust extinction. \citet{marchesini10} found that the inclusion of such a template caused approximately half of the massive galaxies population at $z>3$ to be consistent with a somewhat lower redshift in the range $2<z<3$. For a detailed description of the properties of this template and investigation of its effects on the estimated photometric redshifts we refer the reader to \citet{brammer17}.

The results of the detailed analysis of the spectroscopic data of the brightest $K$-band candidate from \citet{marchesini10}, namely C1-23152, were presented in \citet{marsan15}.
This paper presented the first spectroscopic confirmation of an ultra-massive 
galaxy at redshift $z>3$, along with a detailed investigation of the stellar 
population and structural properties of a progenitor of local most massive 
elliptical galaxies when the universe was less than 2 Gyr old, showing its 
ultra-compact nature, the presence of an obscured powerful AGN, and 
discussing its evolutionary path to $z=0$. We refer the reader to 
\citet{marsan15} for the derived stellar population properties of this source.

\subsection{Emission Line Features and Spectroscopic Redshift}\label{sec-lines}

\begin{deluxetable*}{lccccc}
\centering
\tablecaption{Spectral Line Properties\label{tab-lines}}
\tablehead{\colhead{Feature} & 
           \colhead{$\lambda_{\rm{lab}}$}& \colhead{$z$}& 
           \colhead{FWHM}& \colhead{EW$_{\rm obs}$} & \colhead{$L$} \\ 
  & \colhead{(\AA)} &  & \colhead{(km/s)} & \colhead{(\AA)} & \colhead{($10^{42}$ erg~s$^{-1}$)} } 
\startdata
{\bf C1-19536} & & & & & \\
Ly$\alpha$ & 1215.24 & 3.1544$^{+0.0017}_{-0.0027}$ & 1535.4$^{+1845.6}_{-905.8}$ & 280.4$\pm442.4$ & 11.5$^{+4.9}_{-11.9}$ \\
OIII(Xsh) & 5008.240 & 3.1396$^{+0.0007}_{-0.0008}$ & 896.53$^{+89.67}_{-87.51}$ & 
520.63$^{+142.86}_{-104.16}$ & 26.55$^{+4.49}_{-3.83}$ \\
OIII(NIRSPEC) & 5008.240 & 3.1373$^{+0.0007}_{-0.0008}$ & 871.82$^{+212.09}_{-182.63}$ & 
636.31$^{+254.28}_{-142.78}$ & 17.62$^{+5.93}_{-4.64}$ \\
{\bf C1-15182} & & & & & \\
OIII(Xsh) & 5008.240 & 3.3708$^{+0.0014}_{-0.0021}$ & 734.0$^{+479.4}_{-365.9}$ & 
54.8$^{+2.2}_{-18.3}$ & 16.0$^{+6.8}_{-5.3}$ \\
OIII(NIRSPEC) & 5008.240 & 3.3692$^{+0.0025}_{-0.0021}$ & 767.9$^{+127.7}_{-152.2}$ & 
221.7$^{+108.6}_{-76.3}$ & 7.9$^{+5.9}_{-2.8}$ \\

\enddata
\tablecomments{$\lambda_{lab}$ is the laboratory rest-frame wavelength of the targeted spectral lines; $z$ is the derived redshift; FWHM is the intrinsic velocity width of the spectral lines; EW$_{obs}$ is the observed equivalent widths of the spectral lines; and $L$ is the integrated line luminosity calculated using the adopted systemic redshift $z_{spec}$.} 
\end{deluxetable*}

\begin{figure*}
\epsscale{1}
\plotone{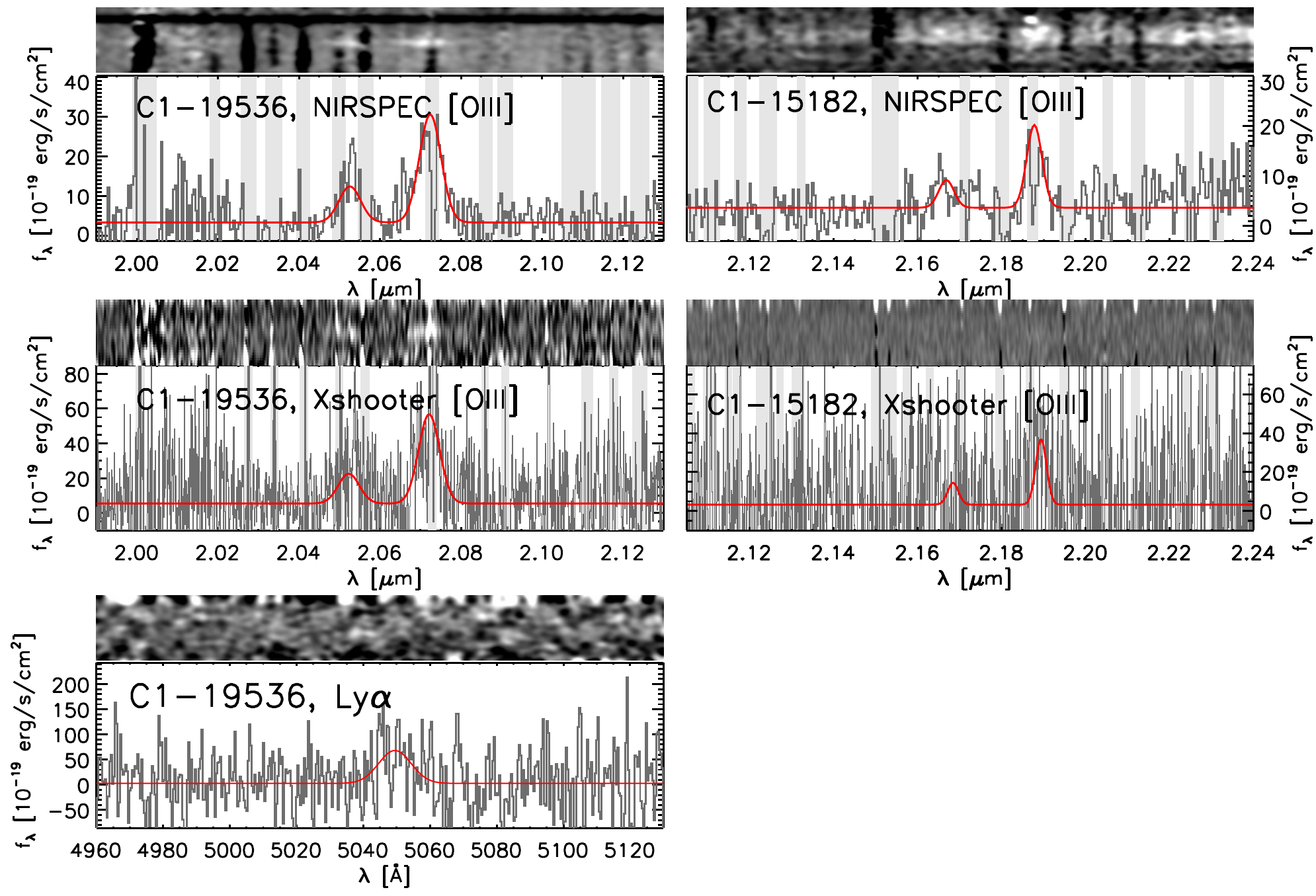}
\caption{The observed one-dimensional and two-dimensional spectra in the
  regions around considered spectral features for C1-19536 and C1-15182.
  Red solid curves indicate the best-fit single Gaussian profiles to the emission lines
  in 1D spectra. The gray shaded regions indicate regions of the spectra
  significantly affected by telluric sky lines. } \label{fig-lines}

\end{figure*}

The extracted one-dimensional spectra were used to measure spectroscopic redshifts ($z_{spec}$) when nebular emission lines were detected.
The nebular emission lines that we set out to measure were primarily 
Ly$\alpha$ and [OIII]$\lambda\lambda$4959,5007. The [OIII] doublet was identified 
in the spectra of C1-19536 and C1-15182, along with a weak Ly$\alpha$ detection in the 
spectrum of C1-19536.  A symmetric Gaussian profile  plus 
a local continuum was used to fit these lines in the extracted one-dimensional 
spectra. Identical redshifts, FWHM and a ratio of 1:3 for the 
amplitudes of the [OIII]4959,5007 doublet were assumed. Figure~\ref{fig-lines} shows the observed one dimensional spectra around the
regions of the considered spectral lines. The regions of the spectra 
significantly affected by skylines, indicated in gray regions, were excluded
from the line profile modeling. The best-fit Gaussian profiles are indicated in red. 

A Monte Carlo approach was used to measure the uncertainties in the centroid, flux, and width of the fitted emission lines. 
For each 1D spectrum, 1000 simulated spectra were created by perturbing 
the flux of the true spectrum at each wavelength by a Gaussian random 
amount with the standard deviation set by the level of the 1$\sigma$ 
error at that specific wavelength. Line measurements were obtained from the simulated spectra 
in the same manner as the actual data. We computed the formal lower and 
upper confidence limits by integrating the probability distribution of 
each parameter (centroid, width, continuum, and emission line flux) from 
the extremes until the integrated probability is equal to 0.1585.
We calculated the fluxes of the observed emission lines by integrating the 
fitted Gaussian profiles, with uncertainties determined using the 1$\sigma$ 
distribution of integrated fluxes from Monte Carlo simulations as described above. The equivalent widths were calculated individually for the observed and simulated spectra by dividing the integrated line flux by the continuum of the fit, with uncertainties derived as above.

The [OIII]$\lambda\lambda$4009,5007 doublet emission is detected in both NIRSPEC and Xshooter 1D spectra for C1-19536 and C1-15182, however, the lower $S/N$ of the Xshooter spectrum is evident in the panels of Figure~\ref{fig-lines}. This is true especially for C1-15182, where a strong continuum is not detected when fitting the emission lines with the Xshooter spectrum, deeming the calculated equivalent width of [OIII] emission highly uncertain. We therefore fixed the continuum level when fitting the [OIII] lines in the Xshooter spectrum for C1-15182 to the median flux values of the spectral region ($\pm 450${\AA}) around the [OIII] lines and not contaminated by skylines. We used the [OIII] emission line fitting results from the higher S/N NIRSPEC spectrum to analyse the emission line properties of C1-15182 and C1-19536.  

The best-fit redshifts, line velocities corrected for the instrumental 
profile (determined by the width of skylines in each spectra) and the observed equivalent
widths are listed in Table~\ref{tab-lines}, with the quoted uncertainties corresponding to the 1$\sigma$ errors estimated from the Monte Carlo simulations. We estimated the line luminosities by adopting the redshifts
of the [OIII] emission lines as the systemic redshifts, listed in Table~\ref{tab-lines}. [OIII] line luminosities are $L_{\rm [OIII]}=1.76^{+0.6}_{-0.5}\times10^{43}$ erg~s$^{-1}$ and $7.9^{+5.9}_{-2.8}\times10^{42}$ erg~s$^{-1}$ for C1-19536 and C1-15182, respectively, at the high end of [OIII] luminosities found in galaxies harboring AGNs \citep{maschietto08, kuiper11, harrison16}. 

The resulting systemic redshifts for C1-15182 and C1-19536 are $z_{spec}$=3.369$\pm$0.002 and $z_{spec}$=3.1373$\pm$0.0008, respectively, spectroscopically confirming their redshifts to be at $z>3$. Compared to the photometric redshifts derived in \citet{marchesini10} and listed in Table~\ref{tab-obs}, we note that the photometric redshift of C1-19536 is in very good agreement with the spectroscopic redshift. For C1-15182, the photometric redshift is smaller by $\sim4\%$ in $\Delta z/(1+z)$ compared to the spectroscopic redshift, but consistent with it at the $\sim2\sigma$ level. For C1-21316, the photometric redshift is smaller by $\sim6\%$ in $\Delta z/(1+z)$ compared to the spectroscopic redshift provided by \citet{capak10}, but still consistent with it at the $\sim2.4 \sigma$ level. We note however that the corresponding spectrum from which $z_{\rm spec}$ was calculated for C1-21316 has been deemed to be of poor quality \citep{miettinen15}. The top panel of Figure 2 shows the comparison between the photometric redshifts from \citet{marchesini10} and the spectroscopic redshifts or the improved $z_{cont}$. The [OIII] line widths listed in Table~\ref{tab-lines} are $\approx$ 768$^{+123}_{-152}$ km s$^{-1}$ and 871$^{+212}_{-183}$ km s$^{-1}$ for C1-15182 and C1-19536, respectively, typical of AGNs of similar L$_{[OIII]}$ at low-$z$ \citep{hao05} and intermediate-$z$ \citep{harrison16}.

\subsection{Improved Photometric Redshifts}\label{sec-zcont}

\begin{figure}[!h]
\epsscale{1.2}
\plotone{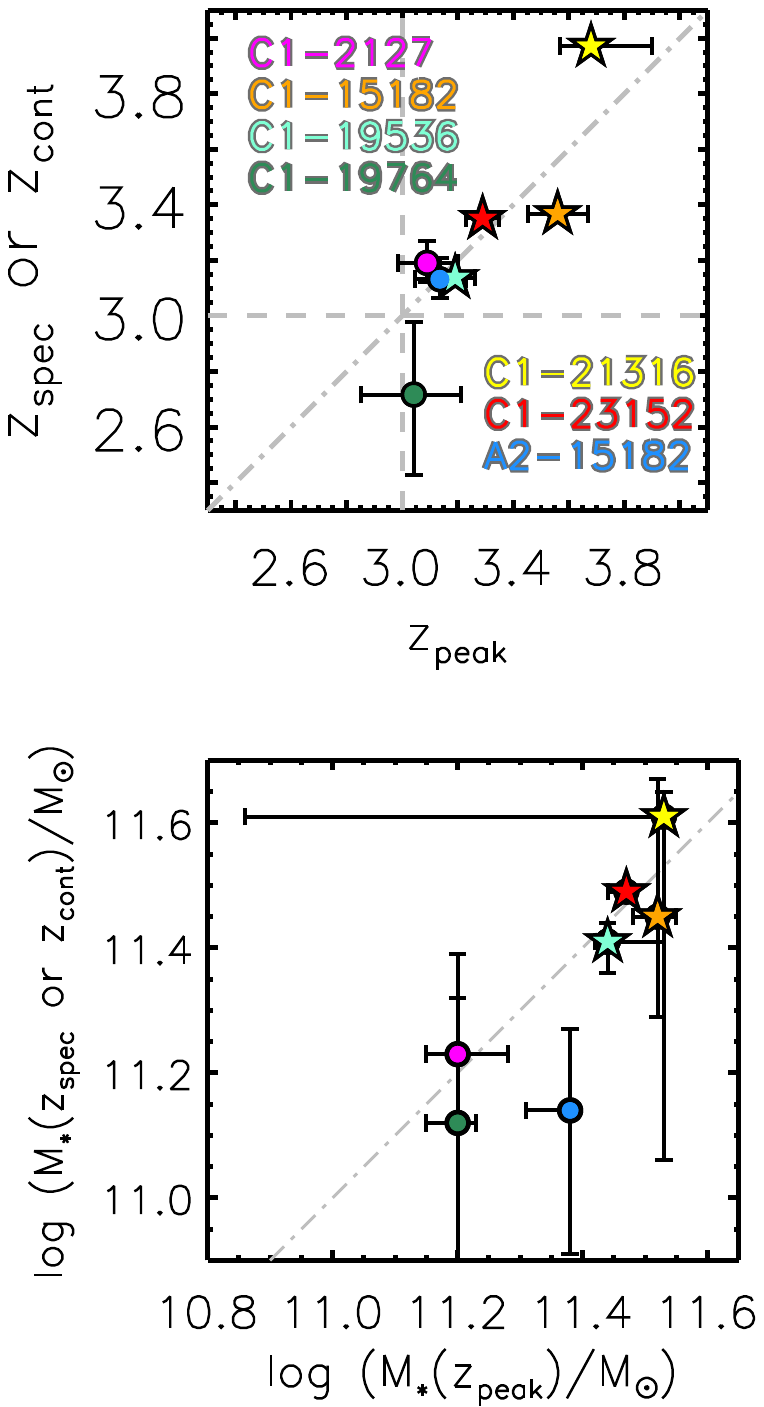}
\caption{Comparison of the best fit stellar masses and redshifts for sources
  with continuum detections and spectroscopic confirmation (C1-23152, {\it red}; C1-19536, {\it aquamarine}; C1-15182, {\it orange}; C1-21316, {\it yellow}; C1-2127, {\it magenta}; C1-19764, {\it green} and A2-15753, {\it blue}). $z_{\rm peak}$ and
  log $M_{*}(z_{\rm peak})$ are from the NMBS catalog v4.4 \citep{marchesini10}. Stars indicate the spectroscopic redshifts when available. Error bars indicate 1$\sigma$ limits output from EAZY \citep{brammer08} and FAST. Gray dot-dashed line is the 1-1 relation for redshifts and steller mass. \label{fig-oldnewpar}}
\end{figure}

\begin{figure*}
\epsscale{0.9}
\plotone{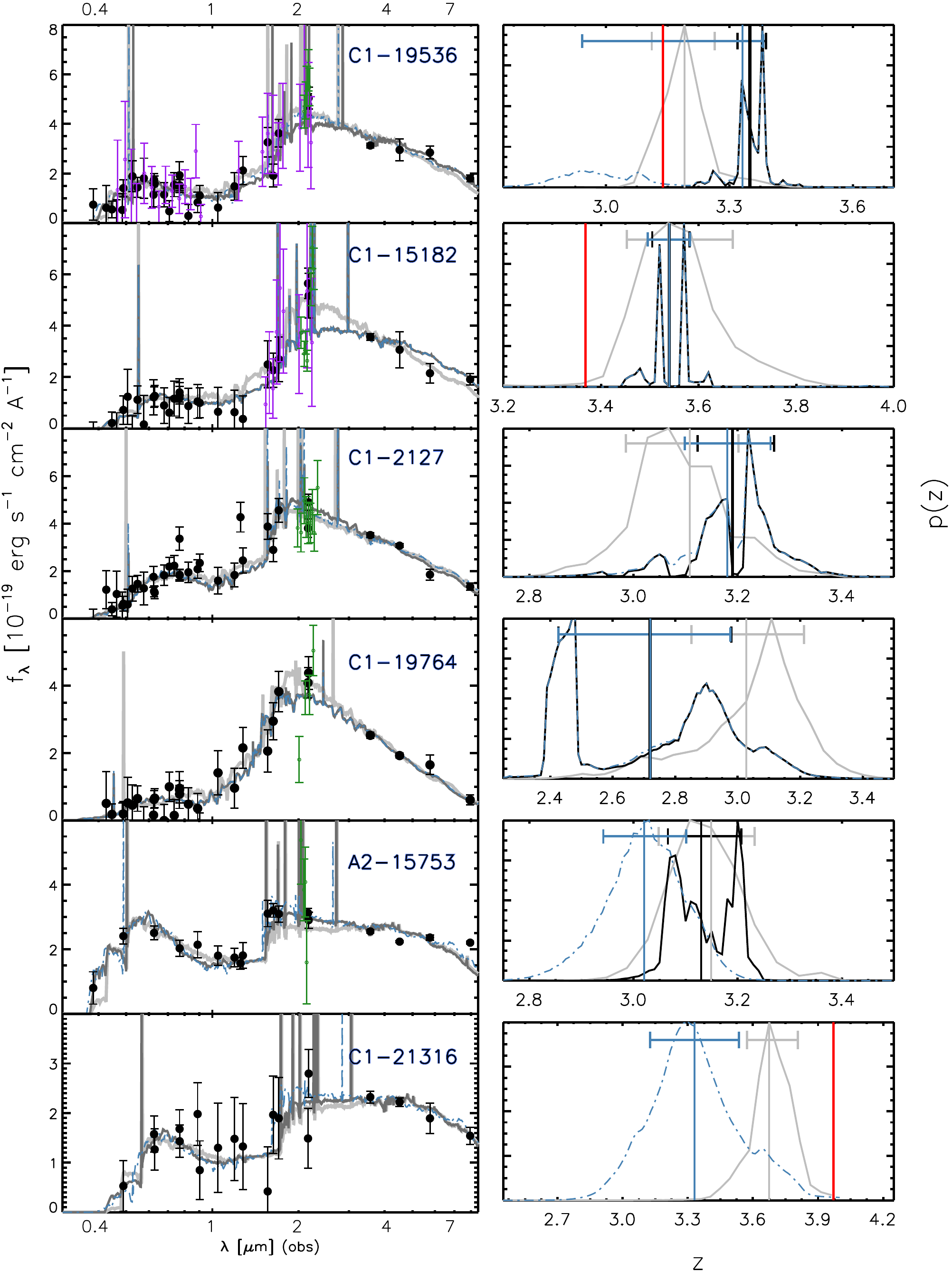}
\caption{{\it Left:} Observed SEDs from the combination of the medium- and broad-band NMBS photometry (black filled circles) and the binned spectra (NIRSPEC, green; X-shooter, purple). The gray curves represent the best-fit EAZY templates using the default EAZY template set adopted in this work (light gray when modeling only the medium- and broad-band photometry; dark gray when also including the 1D binned spectra). The blue dot-dashed curves represent the best-fit EAZY template when the additional old and dusty template is included in the estimate of the photometric redshifts. {\it Right:} The EAZY redshift probability distributions. The gray curves represent the distribution calculated using only medium- and broad-band photometry (NMBS v4.4) whereas the black curves represent the resulting redshift distribution including the 1D binned spectra in EAZY. The blue dot-dashed curves show the resulting redshift probability distributions when the old and dusty SED is included in the template set of EAZY to estimate the photometric redshifts. Vertical lines indicate the $z_{peak}$ output by EAZY for different template sets with 1$\sigma$ uncertainties. Red vertical lines indicate spectroscopic redshifts when available. \label{fig-eazysed}}
\end{figure*}

When no emission lines are detected, we used the 1-D binned spectra combined with the full
FUV-8 $\mu$m broad- and medium-band photometry from the NMBS to determine more accurate 
photometric redshifts ($z_{cont}$). Following \citet{muzzin13a}, the default template set used in this work consists of nine templates: the six templates taken from the optimized template set of EAZY, but augmented with emission lines; a template of a 12.5 Gyr old single stellar population constructed using the \citet{maraston05} models; a 1 Gyr old post-starburst template; and a slightly dust-reddened Lyman break template. The SEDs of the galaxies in our sample are shown in Figure~\ref{fig-eazysed} (left panels), with the EAZY redshift probability functions plotted on the right panels along with the spectroscopic redshift (when available; red), the $z_{peak}$ from \citet{marchesini10} (gray), and the $z_{cont}$ (black).

\citet{marchesini10} found that up to $\sim50\%$ of the massive $3< z < 4$ galaxy sample could be contaminated by a previously unrecognized population of massive, old, and very dusty galaxies at $z<3$. As in \citet{marchesini10}, we considered the case of adding an additional template representing an old (1 Gyr; $\tau = 100$ Myr) and very dusty ($A_{\rm V}=3$ mag) galaxy and repeating the redshift analysis. The resulting redshift probability distributions and best fit $z_{cont}$ are overplotted in the right panels of Figure~\ref{fig-eazysed} (dot-dashed blue). The resulting photometric redshift estimates of only two galaxies in our sample (A2-15753 and C1-21316) are influenced by the addition of the old and dusty template in the analysis, yet both estimates still formally lie at $z>3$. Only the photometric redshift of C1-19764, not included in the \citet{marchesini10} sample, moves below $z=3$ when the old and dusty template is including in EAZY, with a formal solution of $z_{\rm cont}=2.72^{+0.26}_{-0.29}$.

We note that the spectroscopic redshifts of the galaxies in our sample tend to be different from the photometric redshifts at more than $1\sigma$. Specifically, we can see from the right panels of Figure~\ref{fig-eazysed} that EAZY overestimates the redshift of 2/3 of our targets by $\sim$0.17 and $\sim$0.21 for C1-15182 and C1-19536, corresponding to $\sim4\%$ and $\sim5\%$ in $\Delta z / (1+z)$, respectively.

\subsection{SED Modeling and Stellar Population Properties} \label{sec-sedmodel}

\begin{figure*}
\epsscale{1.2}
\plotone{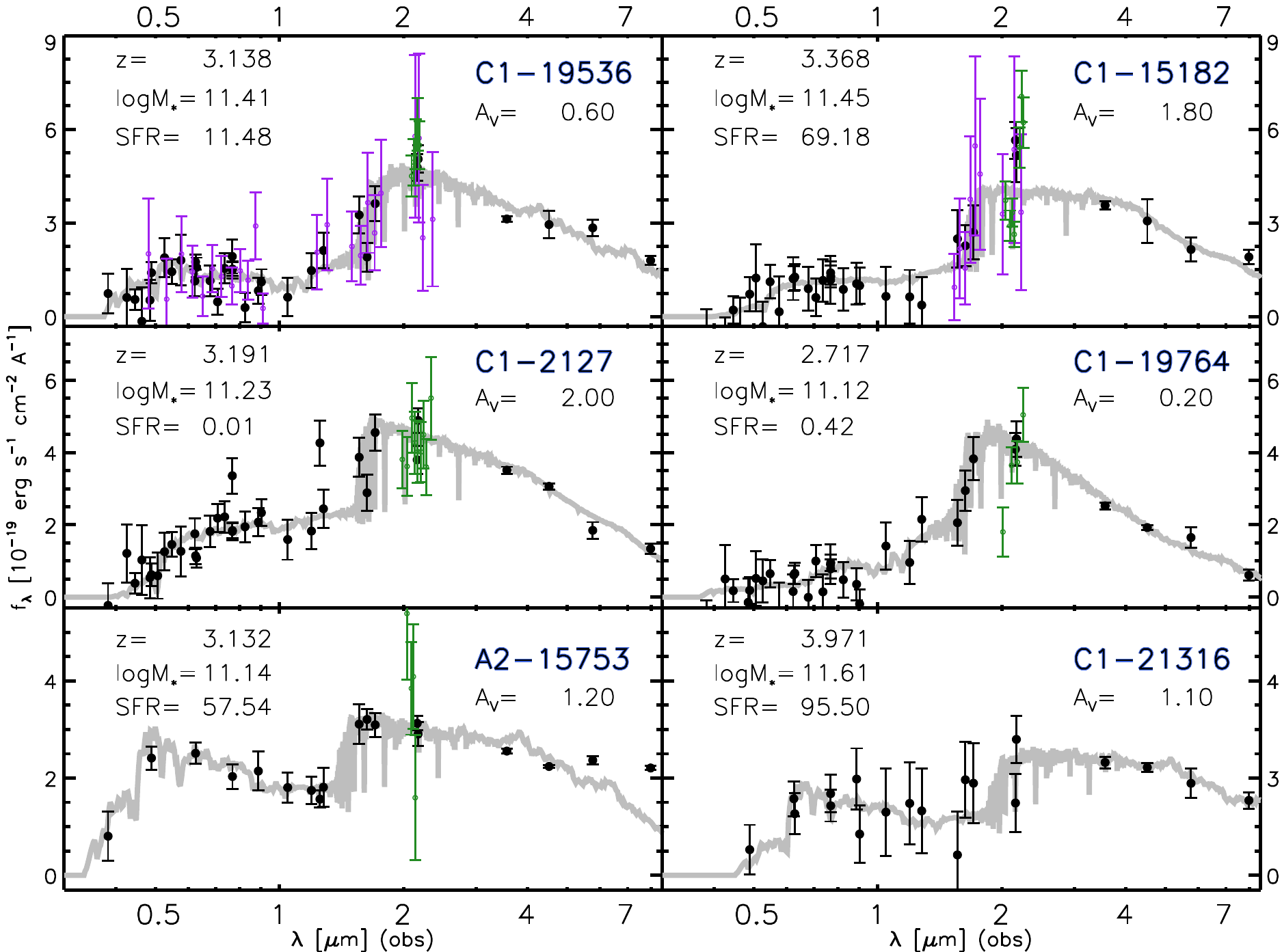}
\caption{{\it Left:} Observed SEDs from the combination of the medium- and broad-band NMBS photometry (black filled circles) and the binned spectra (NIRSPEC in green; X-shooter in purple). The gray curves represent the best-fit models adopting the BC03 population model, with an exponentially declining SFH and free metallicity. Best-fit stellar population properties and redshifts are indicated in each panel. \label{fig-fastsed}}
\end{figure*}

In order to robustly constrain the stellar population parameters, all spectra 
and photometry must be corrected for contamination from nebular emission lines. 
For sources with detected nebular emission lines, we corrected the observed photometry
for emission line contamination by comparing the observed-frame equivalent width of each
emission line to the bandwidth of the corresponding filter. We find that the contribution
due to [OIII] emission is $\sim$15\% and $\sim$5\% for C1-19536 and C1-15182, respectively,
for the NMBS $K$ and $K_{\rm S}$ filters. Due to the highly uncertain equivalent width calculated for the Ly$\alpha$ emission line of C1-19536, we chose to remove the contaminated filter (Subaru, IA505) when proceeding with the SED fittings. 

We estimated the stellar population properties by fitting the binned X-shooter and  NIRSPEC  spectra 
in combination with the broad-band and medium-band photometry with stellar population synthesis (SPS) models. 
We used FAST (Fitting and Assessment of Synthetic Templates; \citealt{kriek09}) to model and fit a full 
grid in metallicity, dust content, age, and star formation timescale.
Stellar population synthesis models of \citet{bc03} were adopted assuming a \citet{kroupa01} IMF, 
an exponentially declining SFH and the \citet{calzetti00} extinction law.
The age of the stellar populations fit ranged between 10 Myr and the maximum age of the universe at the 
redshift of sources with a step size of 0.1 dex. We adopted a grid for $\tau$ between 3 Myr and 10 
Gyr in steps of 0.10 dex and allowed the dust attenuation (A$_{V}$) to range from 0 to 7 mag with step size of 0.01 mag. We initially modeled the observed SED with the metallicity as a 
free parameter ($Z$ = 0.004, 0.08, 0.02, 0.05) and repeated the modeling by treating the metallicity as a systematic uncertainty by fixing it to solar metallicity (Z$_{\odot}$). Figure~\ref{fig-fastsed} shows the observed SEDs of our targets. The results of the SED modeling and the corresponding 1$\sigma$ errors are listed in Table~\ref{tab-sed}. The bottom panel of Figure~\ref{fig-oldnewpar} shows the comparison between the stellar masses from \citet{marchesini10} and the stellar masses derived from the improved redshifts (either $z_{spec}$ or $z_{cont}$) and better sampled SEDs.

As shown in the lower panel of Figure~\ref{fig-oldnewpar}, the stellar masses obtained when adopting the improved redshifts and better sampled SEDs
are consistent with those based only on the NMBS photometry \citep{marchesini10}, spectroscopically confirming the existence of ultra-massive (i.e., log(M$_*/$M$_{\odot}$)>11.4) galaxies at $z>3$, when the universe was younger than 2 Gyr. We note that although error bars on both axes in the panels of Figure~\ref{fig-oldnewpar} correspond to 1$\sigma$ limits, the stellar population parameter space over which the SEDs are modeled are not identical (the observed SEDs are modeled in this work with a finer grid in parameter space than \citealt{marchesini10}, resulting in larger 1$\sigma$ errors). For C1-19536 and C1-15182, which have detected nebular emission lines, we find that removing the emission line contamination to the observed SEDs  decreases the stellar masses by $\sim0.1$ dex (still consistent with being above the mass completeness threshold in \citealt{marchesini10}). The stellar mass of A2-15753 is affected the most when modeling the observed SED including the binned 1D spectra, resulting in a stellar mass of log($M_*/M_\odot$)=11.14$^{+0.13}_{-0.23}$, i.e., $\sim$0.25 dex smaller than what was derived in \citet{marchesini10}.

The stellar mass of C1-2127 and C1-19764, not part of the sample of ultra-massive galaxies at $z>3$ from \citet{marchesini10}, remain effectively the same, i.e., log($M_*/M_{\odot}$)=11.23$^{+0.12}_{-0.44}$ and 11.12$^{+0.20}_{-0.24}$. C1-21316 is the only source with a larger stellar mass when remodeling with improved redshift. This is not surprising as the updated spectroscopic redshift of the source is $\Delta z\sim$ 0.3 greater than the $z_{phot}$ quoted in \citet{marchesini10}. 

The time elapsed since the onset of star formation for our targets range from $\sim$400 Myr to $\sim$ 1.25 Gyr, consistent with having formation redshifts of $z_{\rm form}\sim$ 5-9. For roughly half of the sample the best-fit SFR-weighted mean ages $<t>_{\rm SFR}$ (as defined in \citealt{forsterschreiber04}) are $\approx 800$~Myr. Two of the massive galaxies (C1-19764, with $z_{\rm cont}<3$ and C1-2127) in our sample have best-fit SFR estimates from SED modeling consistent with low SFR ($<1$ M$_{\odot}$yr$^{-1}$), while the remaining galaxies have SFRs on the order of few tens to hundreds solar masses per year. We stress that the estimated SFRs, stellar ages, and dust extinction values have large uncertainties. 

\begin{deluxetable*}{lcccccccc}
\centering
\tablecaption{Best Fit Stellar Population Parameters}
\tablehead{\colhead{ID}  & \colhead{$z^{a}$} & \colhead{log$\tau$} & \colhead{Metallicity} &
  \colhead{log($Age$)} & \colhead{A$_{\rm V}$} & \colhead{log($M_{*}$)} & 
  \colhead{SFR} & \colhead{$\chi^2$} \\ 
    & & \colhead{(yr)} &  & \colhead{(yr)} & \colhead{(mag)} &
  \colhead{M$_{\odot}$} & \colhead{M$_{\odot}$yr$^{-1}$} & \\}
\startdata

{\bf C1-2127} & 3.19$^{+0.08}_{-0.07}$ & 6.70$^{+3.30}_{-0.20}(^{+3.30}_{-0.20})$ & 0.008$^{+0.042}_{-0.004}(^{+0.042}_{-0.004})$ & 7.90$^{+0.82}_{-0.90}(^{+1.20}_{-0.90})$ & 2.00$^{+1.10}_{-1.10}(^{+1.18}_{-1.70})$ & 11.23$^{+0.16}_{-0.44}(^{+0.26}_{-0.54})$ & 0.0067$^{+7244.}_{-0.006}(^{+8128.}_{-0.006})$ & 1.12  \\

 & & 8.00$^{+0.24}_{0.000}(^{+2.00}_{0.000})$ & 0.02 & 8.60$^{+0.19}_{-0.18}(^{+0.35}_{-1.60})$ & 1.40$^{+0.37}_{-0.47}(^{+1.60}_{-0.90})$ & 11.35$^{+0.06}_{-0.04}(^{+0.14}_{-0.60})$ & 53.7$^{+69.3}_{-35.9}(^{+6402.}_{-49.53})$ & 1.17  \\

\\
\hline
\\ 
{\bf C1-15182} & 3.3703$^{+0.0013}_{-0.0011}$ & 8.00$^{+0.92}_{-1.50}(^{+2.00}_{-1.50})$ & 0.004$^{+0.046}_{0.000}(^{+0.046}_{0.000})$ & 8.60$^{+0.60}_{-0.79}(^{+0.60}_{-1.60})$ & 1.80$^{+0.59}_{-0.73}(^{+1.52}_{-1.26})$ & 11.45$^{+0.22}_{-0.16}(^{+0.28}_{-0.52})$ & 69.2$^{+367.3}_{-69.18}(^{+10895}_{-69.2})$ & 1.86  \\

  & & 8.30$^{+0.33}_{-1.14}(^{+0.76}_{-1.80})$ & 0.02 & 8.80$^{+0.39}_{-0.71}(^{+0.40}_{-1.03})$ & 1.60$^{+0.50}_{-0.61}(^{+0.72}_{-1.04})$ & 11.56$^{+0.09}_{-0.16}(^{+0.18}_{-0.29})$ & 102.3$^{+137.5}_{-96.3}(^{+398.8}_{-102.3})$ & 1.85  \\
    
\\
\hline
\\
{\bf C1-19536} & 3.1385$^{+0.0005}_{-0.0005}$ & 8.30$^{+0.14}_{-0.10}(^{+0.62}_{-0.28})$ & 0.05$^{+0.000}_{-0.014}(^{+0.000}_{-0.046})$ & 9.00$^{+0.14}_{-0.10}(^{+0.30}_{-0.36})$ & 0.60$^{+0.10}_{-0.15}(^{+1.13}_{-0.31})$ & 11.41$^{+0.03}_{-0.05}(^{+0.17}_{-0.15})$ & 11.5$^{+1.4}_{-1.5}(^{+117.3}_{-5.7})$ & 1.35  \\

  & & 8.50$^{+0.15}_{-0.11}(^{+0.30}_{-0.34})$ & 0.02 & 9.10$^{+0.20}_{-0.10}(^{+0.20}_{-0.37})$ & 0.90$^{+0.10}_{-0.51}(^{+0.72}_{-0.61})$ & 11.45$^{+0.09}_{-0.10}(^{+0.18}_{-0.15})$ & 22.9$^{+4.0}_{-15.7}(^{+89.3}_{-16.6})$ & 1.44  \\

\\
\hline
\\
{\bf C1-21316} & 3.971 & 8.70$^{+1.30}_{-2.20}(^{+1.30}_{-2.20})$ & 0.05$^{+0.00}_{-0.046}(^{+0.00}_{-0.046})$ & 9.10$^{+0.00}_{-1.66}(^{+0.00}_{-2.10})$ & 1.10$^{+1.21}_{-0.51}(^{+1.84}_{-0.91})$ & 11.61$^{+0.04}_{-0.55}(^{+0.07}_{-0.94})$ & 95.5$^{+795.7}_{-95.3}(^{+9024.}_{-95.5})$ & 0.64  \\

& & 9.30$^{+0.70}_{-2.46}(^{+0.70}_{-2.80})$ & 0.02 & 9.10$^{+0.00}_{-1.51}(^{+0.00}_{-2.10})$ & 1.60$^{+0.74}_{-0.70}(^{+1.33}_{-1.02})$ & 11.58$^{+0.06}_{-0.52}(^{+0.09}_{-0.91})$ & 281.8$^{+1416.}_{-230.5}(^{+7846.}_{-281.8})$ & 0.68  \\
\\
\hline
\\
{\bf C1-19764} & 2.72$^{+0.26}_{-0.29}$ & 8.20$^{+0.42}_{-1.70}(^{+0.81}_{-1.70})$ & 0.008$^{+0.042}_{-0.004}(^{+0.042}_{-0.004})$ & 9.10$^{+0.30}_{-0.54}(^{+0.30}_{-0.83})$ & 0.20$^{+1.13}_{-0.20}(^{+2.19}_{-0.20})$ & 11.12$^{+0.20}_{-0.24}(^{+0.26}_{-0.39})$ & 0.4$^{+3.8}_{-0.4}(^{+28.4}_{-0.4})$ & 0.81  \\

 &  & 8.10$^{+0.30}_{-1.60}(^{+0.79}_{-1.60})$ & 0.02 & 9.00$^{+0.25}_{-0.46}(^{+0.40}_{-0.81})$ & 0.20$^{+1.02}_{-0.20}(^{+2.12}_{-0.20})$ & 11.20$^{+0.10}_{-0.13}(^{+0.18}_{-0.40})$ & 0.6$^{+3.7}_{-0.6}(^{+28.9}_{-0.6})$ & 0.83  \\

\\
\hline
\\ 

{\bf A2-15753} & 3.13$^{+0.07}_{-0.07}$& 8.90$^{+1.10}_{-0.51}(^{+1.10}_{-2.40})$ & 0.05$^{+0.00}_{-0.033}(^{+0.00}_{-0.046})$ & 9.10$^{+0.30}_{-0.43}(^{+0.30}_{-1.80})$ & 1.20$^{+0.40}_{-0.39}(^{+1.36}_{-0.80})$ & 11.14$^{+0.13}_{-0.23}(^{+0.26}_{-0.73})$ & 57.5$^{+74.3}_{-32.4}(^{+965.7}_{-57.5})$ & 0.91  \\
  
  & & 10.0$^{+0.00}_{-0.98}(^{+0.00}_{-3.50})$ & 0.02 & 9.40$^{+0.00}_{-0.41}(^{+0.00}_{-1.79})$ & 1.40$^{+0.31}_{-0.27}(^{+0.82}_{-0.68})$ & 11.17$^{+0.13}_{-0.15}(^{+0.23}_{-0.60})$ & 69.2$^{+62.6}_{-28.4}(^{+328.9}_{-69.2})$ & 1.05  \\

\enddata

\tablecomments{Estimated stellar population parameters from the modeling of the binned UV-to-NIR spectra in combination with the broad- and medium-bandwidth UV-to-8$\mu$m photometry from NMBS derived assuming the \citet{bc03} stellar population synthesis models and an exponentially declining SFH. Spectroscopic redshifts are listed when available, otherwise best-fit EAZY outputs are listed. Quoted errors are 1$\sigma$ confidence intervals output by FAST and EAZY. The values in parantheses correspond to 3$\sigma$ confidence intervals. A \citet{kroupa01} IMF and a \citet{calzetti00} extinction law is assumed in all cases.} \label{tab-sed}
\end{deluxetable*}

\begin{figure}
  \epsscale{1.15}
  \plotone{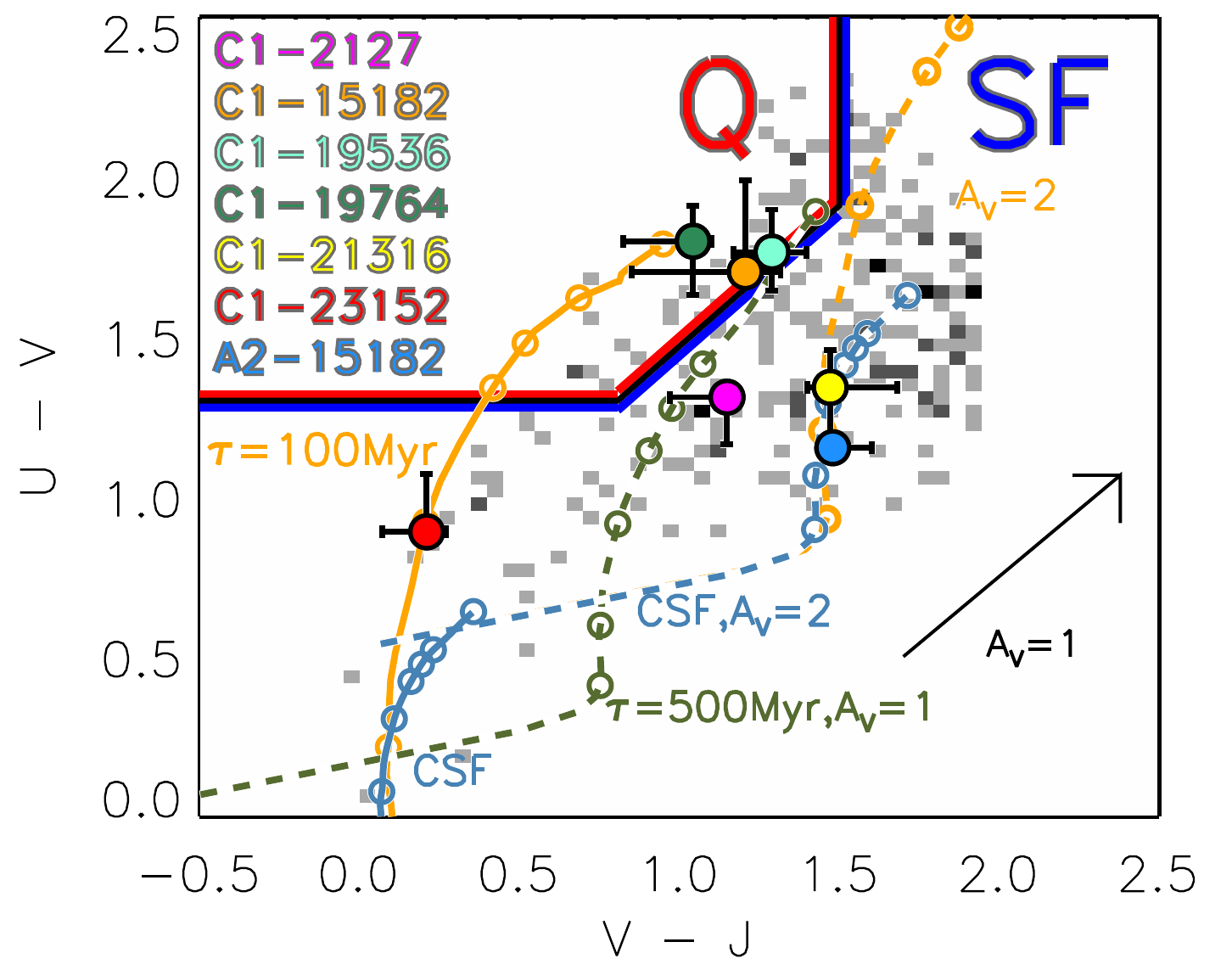}
  \caption{Rest-frame $U - V$ vs. $V - J$ color-color diagram. The grayscale representation indicates the distributions of $3.0< z <4.0$ galaxies above the $95\%$ stellar mass completeness limits from the $K_{\rm S}$-selected UltraVISTA catalog \citep{muzzin13a}. The cuts used to separate star-forming from quiescent galaxies from \citet{muzzin13b} are shown as the solid black lines. Galaxies are indicated with the identical color scheme as in Figure~\ref{fig-oldnewpar}. Color evolution tracks of \citet{bc03} models are also shown: an exponentially declining SFHs with $\tau = 100$ Myr in orange ($A_{\rm}= 0$ mag, solid;  $A_{\rm V}= 2$ mag, dashed) and $\tau =500$ Myr in green with $A_{\rm V}=1$, and constant SFHs in blue ($A_{\rm}= 0$ mag, solid;  $A_{\rm V}= 2$ mag, dashed). The dust vector indicates an extinction of $A_{\rm V}=$ 1 mag for a \citet{calzetti00} extinction curve.\label{fig-uvj}}
  \end{figure}

Figure~\ref{fig-uvj} shows the rest-frame $U-V$ versus $V-J$ diagram (hereafter, $UVJ$ diagram) with the rest-frame colors of our targets (color scheme identical to Figure~\ref{fig-oldnewpar}) overlayed on the distribution of the rest-frame colors of all galaxies at $3.0 <z< 4.0$ from the $K_S$-selected UltraVISTA catalog of \citet{muzzin13a} (grayscale representation). The rest-frame $UVJ$ diagram is a powerful diagnostic to separate star-forming and quiescent galaxy populations in color-color space \citep{labbe05, wuyts07, brammer11, whitaker11, muzzin13b}.  We calculated the rest frame $U-V$ and $V-J$ colors of the targets using EAZY \citep{brammer08}. A Monte Carlo approach was used to measure the uncertainties in the $U - V$ and $V - J$ colors. Specifically, 1000 photometry catalogs were created by perturbing each flux by a Gaussian random number with the standard deviation set by the level of each flux error. The simulated catalogs were each fit with EAZY separately, and the formal upper and lower limits were obtained in a similar manner as for the emission line fits. The rest-frame colors and 1$\sigma$ confidence limits for the galaxies in our sample are plotted in Figure~\ref{fig-uvj} in solid colored circles (color scheme identical to Figure~\ref{fig-oldnewpar}). Figure~\ref{fig-uvj} also shows color-color evolution tracks of \citet{bc03} models stellar population synthesis models for different SFHs and dust extinction values. The orange solid and dashed tracks represent the evolution of an exponentially declining SFH with $\tau=$ 100 Myr and $A_{\rm V}=$0 and 2 mag, respectively. The blue solid and dashed tracks represent the evolution for a constant SFH with $A_{\rm V}=$0 and 2 mag, respectively. The evolution of an exponentially declining SFH with $\tau=$500 Myr and $A_{\rm V}=1$ is indicated by the dashed green track. 

Based on rest-frame colors, only one (C1-19764, only galaxy with best-fit $z_{cont}< 3$) of the targets falls firmly in the quiescent region. Two galaxies with confirmed $z_{\rm spec}$, C1-19536 and C1-15182, that are also hosts to powerful AGNs (based on the X-ray detections see Section~\ref{sec-xrad}), are on the border of the quescient - star forming galaxy rest-frame color separation. The quenched Vega galaxy (C1-23152, \citealt{marsan15}), has rest-frame colors consistent with transitioning into the quiescent region. The positions of the remaining three targets (C1-2127, C1-21316 and A2-15753) indicate that they are dusty, star-forming galaxies, in line with the derived stellar population parameters based on SED modeling. 

\subsection{Active Galactic Nuclei Content}

\subsubsection{Infrared Spectral Energy Distribution}\label{sec-irsed}

\begin{deluxetable*}{lccccc}
\centering
\tablecaption{FIR Flux Density Detections and Limits \label{tab-fir}}
\tablehead{ \colhead{ID} & \colhead{$S_{\nu}$(100$\mu$m)} & \colhead{$S_{\nu}$(160$\mu$m)} & \colhead{$S_{\nu}$(250$\mu$m)} & \colhead{$S_{\nu}$(350$\mu$m)} & \colhead{$S_{\nu}$(500$\mu$m)} \\
  & \colhead{(mJy)} & \colhead{(mJy)} & \colhead{(mJy)} & \colhead{(mJy)} & \colhead{(mJy)} }
\startdata

C1-2127 & (1.8) & (3.4) & 13.4$^{+2.2}_{-2.3}$ & 22.2$^{+4.7}_{-4.7}$& (10.9) \\
C1-15182 & (1.7) & (3.4) & (3.6)& (2.1) & (2.9) \\
C1-19536 & (1.7)& (3.4) & (2.7) & (3.6) & (5.1) \\
C1-19764 & (1.7) & (3.4) & (2.7) & (3.6) & (5.1) \\
C1-21316 & (2.0) & (3.4) & 29.1$^{+2.5}_{-3.2}$ & 38.3$^{+3.1}_{-3.4}$ & 31.6$^{+6.6}_{-29.2}$ \\
A2-15753 & (1.2) & (2.5) & (3.6) & (5.9) & (5.6) \\

\enddata
\tablecomments{The {\it Herschel} PACS 100, 160$\mu$m, SPIRE 250, 350, 500$\mu$m detections and limits. Photometric points with detections are listed with the corresponding 1$\sigma$ errors, whereas non-detections are represented with the 1$\sigma$ error limits in paranthesis.}
\end{deluxetable*}

\begin{deluxetable*}{lccccccccc}
\centering
\tablecaption{{\it Spitzer}-24$\mu$m Fluxes and the Derived $L_{\rm IR}$ and SFR}
\tablehead{ \colhead{ID} & \colhead{$S_{24}$} & \colhead{$L_{\rm IR}^a$}& \colhead{SFR$_{\rm IR}^a$} &  \colhead{$L_{\rm IR}^b$} &\colhead{SFR$_{\rm IR}^b$} & \colhead{$f(\rm AGN)_{bol}^c$} & \colhead{$f(\rm AGN)_{MIR}^c$} &\colhead{$L_{\rm UV}$}& \colhead{SFR$_{\rm UV}$} \\
  & \colhead{($\mu$Jy)} & \colhead{(10$^{12}$ L$_{\odot}$)} & \colhead{(M$_{\odot}$ yr$^{-1}$)} & \colhead{(10$^{12}$ L$_{\odot}$)}  & \colhead{(M$_{\odot}$ yr$^{-1}$)} & \colhead{$\%$}
  & \colhead{$\%$} &\colhead{(10$^{9}$ L$_{\odot}$)} & \colhead{(M$_{\odot}$ yr$^{-1}$)} } 
\startdata
C1-2127  & 110.6$\pm$10.7 & 9.6$\pm$0.9($^{+2.0}_{-1.5}$)   & 1049$\pm$101($^{+220}_{-168}$) & 7.2$\pm$0.7($^{+0.8}_{-0.8}$) & 761$\pm$291($^{+295}_{-294}$) & 
	2.7$\pm1.0$ & $\sim15$ & 5.3 & 1.7 \\
	
C1-15182 & 78.6$\pm$7.3  & 10.3$\pm$1.0  & 1119$\pm$104  & <~2.6$\pm$0.24 & <~259$\pm$96 &
	>~$9.2\pm 3.2$ & $>50$ & 5.6 & 1.8 \\

C1-19536 & 61.6$\pm$6.7  & 4.7$\pm$0.5    & 512$\pm$56   & <~3.6$\pm$0.4 & <~384$\pm$148 & 
	>$~4\pm 1$ &  $>15$ &4.8 & 1.6 \\ 

C1-21316 & 177.1$\pm$7.8  & 46.0$\pm$1.2  & 4993$\pm$64  & 9.1$\pm$0.4 - 14.9$\pm$0.7  & 900$\pm$325 - 1516$\pm$540 & 
	$4 - 12 $  &$\sim 35-50$ & 5.4 & 1.8 \\

A2-15753 & 165.7$\pm$4.2 & 13.3$\pm$0.6($^{+2.5}_{-2.1}$)   & 1357$\pm$34($^{+276}_{-227}$)  & <~5.3$\pm$0.1($^{+0.3}_{-0.3}$)  & <~533$\pm$188($^{+190}_{-190}$) &
	 >$~10\pm 4$  & $>40$ & 8.5 & 2.7 \\
\enddata
\tablecomments{The $Spitzer$ 24$\mu$m fluxes and derived $L_{\rm IR}$ and SFR$_{\rm IR}$ assuming: $^a$ the mean \citet{dale02} IR template, and $^b$ the best-fit mean \citet{dale14} IR template in agreement with $Herschel$ 3$\sigma$ detection limits; these values are strictly {\it upper limits} to the $L_{\rm IR}$ and SFRs allowed by the observed FIR fluxes and limits; $^{c}$ lists the fraction of AGN luminosity contribution to the total bolometric ($5-1100\mu$m) and MIR ($5-20\mu$m) luminosity (values in paranthesis) for the best-fit mean \citet{dale14} templates; these values are lower limits constrained by FIR fluxes and limits.
For C1-21316, the two values provided in columns 5 to 8 correspond to values derived assuming the two \citet{dale14} templates constrained by either the SCUBA 450~$\mu$m or the $\lambda>$800~$\mu$m fluxes (solid blue curves in Figure~\ref{fig-ir}). 
The errors listed for L$_{IR}$ and SFR$_{\rm IR}$ were computed using just the 24$\mu$m photometric errors (values not in parentheses) and the combination of the 24$\mu$m photometric errors and the photometric redshift errors (values in parentheses).}  \label{tab-ir}
\end{deluxetable*}

\begin{figure*}
\epsscale{1}
\plotone{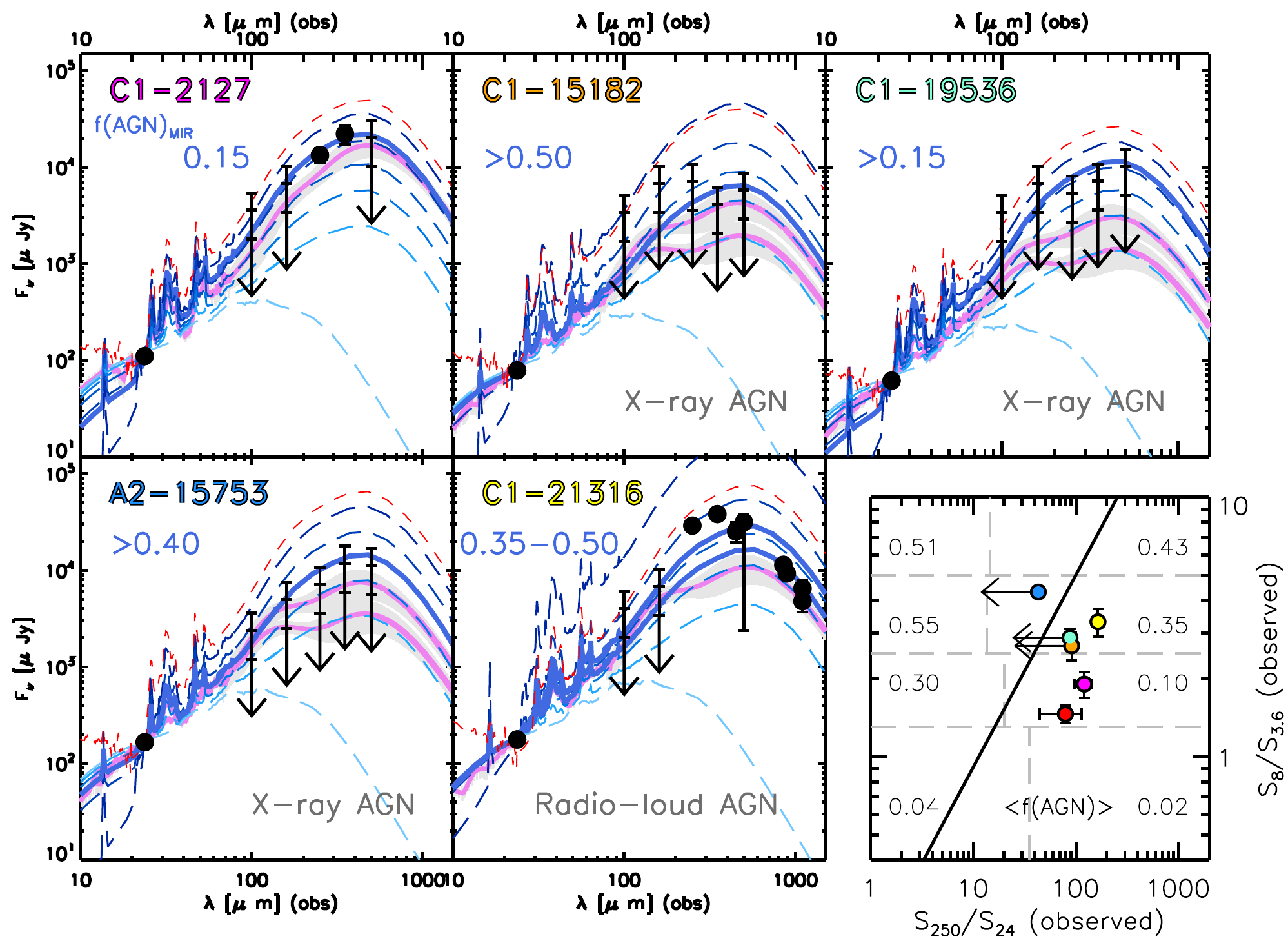}
\caption{FIR SED with {\it Spitzer} MIPS 24$\mu$m, {\it Herschel} PACS 100, 160$\mu$m, SPIRE 250, 350, 500$\mu$m with IR templates overplotted. 
For C1-21316 (AzTEC-5), the plotted photometry also includes the de-boosted fluxes from SMA at 890~$\mu$m \citep{younger07}, JCMT AzTEC at 1.1mm \citep{scott08}, ASTE AzTEC at 1.1mm \citep{aretxaga11}, and JCMT SCUBA2 at 450 and 850$\mu$m \citep{casey13}.
Filled black circles represent observations with detections $S/N>1$. Photometric points with no detection are indicated with their 1, 2, 3$\sigma$ upper limits. The red dashed SED represents the average SED of \citet{dale02} templates used in \citet{marchesini10}. Dashed (blue) templates represent the mean SEDs for $\alpha$ = 1,...,2.5 from the template set of \citet{dale14} for varying degrees of AGN contribution (dark blue: 0$\%$ to light blue: 100$\%$ AGN contribution in uniform 20$\%$ steps). 
The thick sold curve represents the template from \citet{dale14} with the minimum amount of AGN contamination allowed by the IR fluxes or 3$\sigma$ upper limits. The minimum amount of AGN contribution is also specified in blue. Two blue solid curves and AGN values are provided for C1-21316, depending on the IR data used to constrain the AGN contamination (the SCUBA 450~$\mu$m or the fluxes at $\lambda>$800~$\mu$m).
Violet SEDs are the IR color-based templates from \citet{kirkpatrick15}. 
Each panel specifies whether the galaxy is also an X-ray AGN or a radio-loud AGN. {\it Bottom right panel:} Distribution of our sources in IR colorspace using the observed frame colors (colors same as in Fig~\ref{fig-oldnewpar}). The black line represents the empirical separation between the AGN and SFGs defined in \citet{kirkpatrick13}, the gray dashed lines show the dividing lines between color regions from which average templates are calculated in \citet{kirkpatrick15}. The $<f(AGN)_{\rm IR}>$ for each colorspace is indicated in the corresponding region \citep{kirkpatrick15}. } \label{fig-ir}

\end{figure*}

We found that 5/6 of our targets have significant (> 3$\sigma$) MIPS 24$\mu$m detections (four already studied in \citealt{marchesini10}), consistent
with the high fraction of MIPS-detected sources in the sample of IRAC-selected massive galaxies at $z > 3.5$ over GOODS-North \citep{mancini09}. At the redshifts of our targets, the observed 24$\mu$m probes the rest-frame $\sim$4.8-7.8 $\mu$m. Emission at 
these wavelengths can arise from warm/hot dust and polycyclic aromatic hydrocarbons (PAHs; \citealt{draineli07}), heated by dust-enshrouded star formation or AGN. 

Figure~\ref{fig-ir} shows the observed far-infrared (FIR) SEDs of our targets with significant MIPS-24$\mu$m detections. 
We include the {\it Herschel} PACS 100~$\mu$m and 160~$\mu$m photometry from the PACS Evolutionary Probe (PEP) survey DR1 public data release\footnote{\url{http://www.mpe.mpg.de/ir/Research/PEP/DR1}} \citep{lutz11,magnelli13} and {\it Herschel} SPIRE 250~$\mu$m, 350~$\mu$m, and 500~$\mu$m photometry from the Herschel Multi-tiered Extragalactic Survey (HerMES) DR3/4 public data release\footnote{\url{http://hedam.lam.fr/HerMES/index/dr4}} \citep{roseboom10, oliver12, roseboom12, hurley16}. These fluxes are listed in Table~\ref{tab-fir}. For C1-21316 (a.k.a. AzTEC 5), we also included the de-boosted photometry from SMA at 890~$\mu$m ($S_{890\mu\rm  m}=9.3\pm1.3$~mJy; \citealt{younger07}), JCMT AzTEC at 1.1mm ($S_{1.1{\rm mm}}=6.5\pm1.4$~mJy; \citealt{scott08}), ASTE AzTEC at 1.1mm ($S_{1.1\rm{mm}}=4.8\pm1.1$~mJy; \citealt{aretxaga11}), and JCMT SCUBA2 at 450$\mu$m and 850$\mu$m ($S_{450\mu\rm m}=25.35\pm6.04$~mJy and $S_{850\mu\rm m}=11.42\pm1.38$~mJy; \citealt{casey13}). Most of our sources are not detected in any of the {\it Herschel} bands. Only C1-21316, which is a sub-millimeter galaxy and a radio-loud AGN, is detected in SPIRE and at longer wavelengths; C1-2127 is detected in the blue and green SPIRE bands, but not at 500~$\mu$m. No source is detected in PACS despite all being robustly detected in the MIPS 24$\mu$m band. Figure~\ref{fig-ir} shows the IR SEDs of the targeted sources, along with the 1, 2, and 3$\sigma$ upper limits indicated for the bands without detections.

{The 24$\mu$m emission has been widely used in the literature to estimate total infrared luminosities ($L_{\rm IR}=L_{\rm 8-1000\mu m}$; see, e.g., \citealt{papovich07, rigby08}), which can then be transformed into dust-enshrouded SFRs \citep{kennicutt98}. In this section we will show that this approach cannot be blindly adopted in ultra-massive galaxies at $z>3$ given the almost ubiquitous presence of obscured AGN in this population and the resulting AGN contamination of the 24$\mu$m emission. 

First, we derived dust-enshrouded SFRs from the MIPS 24$\mu$m fluxes using the approach presented in \citet{wuyts08}, which has become one of the most adopted approaches in the literature. Specifically, this method uses the mean log$L_{\rm IR,\alpha=1,...,2.5}$ of the infrared SEDs of star-forming galaxies provided by \citet{dale02} to calculate SFR$_{\rm IR}$. This method was validated out to $z\sim3.5$ for lower mass (i.e., $\log{(M_{\star}/M_{\odot})}<11$) star-forming galaxies using Herschel data \citep{wuyts11, tomczak16}.
The calculated total 8-1000$\mu$m rest-frame luminosities and the corresponding SFR$_{IR}$ adapted for a \citet{kroupa01} IMF from \citet{kennicutt98} are listed in Table~\ref{tab-ir}, third and fourth columns. The corresponding errors are also listed, with and without the uncertainties due to random photometric redshift uncertainties when spectroscopic redshifts are not available. As shown in Table~\ref{tab-ir}, the random uncertainties on L$_{\rm IR}$ and SFR$_{\rm IR}$ as derived using the \citet{wuyts08} approach are dominated by the contribution from random photometric redshift errors. We note that a significant scatter of $\sim0.2-0.3$dex was found by \citet{wuyts11} and \citet{tomczak16} between the MIPS 24$\mu$m derived SFRs and the SFRs derived including robust {\it Herschel} PACS detections. Therefore, we stress that the total error budget of L$_{\rm IR}$ and SFR$_{\rm IR}$ listed in Table~\ref{tab-ir} as derived using the \citet{wuyts08} approach is dominated by this scatter and can be as large as a factor of a few.
When the \citet{wuyts08} approach is adopted, the $L_{\rm IR}$ of our targets range from $\sim 5 \times 10^{12}L_{\odot}$ to $\sim 5 \times 10^{13}L_{\odot}$, typical of Ultra Luminous IR galaxies (ULIRGs) and Hyper Luminous IR galaxies (HLIRGs) \citep{murphy11}, corresponding to SFR $\sim 500-5000$ M$_{\odot}$yr$^{-1}$ of obscured star-formation. 
We used the rest-frame 2800 {\AA}  luminosity ($\nu L_{2800{\AA}}$), determined via the best-fit templates in EAZY (using the same methodology for deriving rest-frame colors described in \citealt{brammer11}), as a proxy for $L_{\rm UV}$, reflecting the contribution of unobscured star formation. We used the approach detailed in \citet{bell05} to convert $\nu L_{2800{\AA}}$ to SFR$_{\rm UV}$. The calculated L$_{\rm UV}$ and SFR$_{\rm UV}$ are listed in Table~\ref{tab-ir}.

The SFRs derived by assuming that all the flux at the observed 24 $\mu$m is associated with dust-enshrouded star formation are $\sim 2-3$ orders of magnitude larger than the SFRs estimated from SED modeling (although values for C1-2127 and A2-15753 are consistent within 1$\sigma$ uncertainties). 
We can see in Figure~\ref{fig-ir} that the median pure-SF IR SED template derived from the library of \citet{dale02}, and adopted in the \citet{wuyts08} approach, is not an adequate fit to the observed detections and limits. For each object, we overplot the median \citet{dale02} template scaled to the observed MIPS-24$\mu$m detections as dashed red curves. It can be seen that a model in which the IR emission is due only to obscured star formation cannot reproduce the observed IR SEDs. Specifically, the detections and upper limits in {\it Herschel} bands put strong constraints on the overall shape of the IR SEDs, rejecting the scenario in which the observed IR properties are due entirely to obscured star-formation activity, and pointing to the presence of a component with warmer dust temperature (arguably originating from the dusty torus of the AGN) not present in the average template from \citet{dale02}. This result was already presented in \citet{marsan15} for the Vega galaxy at $z=3.35$, and we now show that it appears to be a common property of ultra-massive galaxies at $z > 3$. 

To quantitatively assess the amount of AGN contribution to the IR SEDs, we used the IR template set of \citet{dale14} which includes fractional AGN contributions to the mid-infrared (MIR) radiation. We calculated the mean log($L_{\rm IR}$) for $\alpha=$ 1, ..., 2.5 from this template set for varying AGN contributions. For visual ease in comparison, we only plot the templates with AGN contributions of 0$\%$, 20$\%$, 40$\%$, 60$\%$, 80$\%$ and 100$\%$ scaled to the observed MIPS 24$\mu$m photometry (dashed blue curves in Figure~\ref{fig-ir}). The \citet{dale14} templates that minimize the $\chi^2$ to detections and are consistent with upper limits for each target are plotted as solid blue curves, with f(AGN)$_{\rm MIR}$ indicated in each panel. We note that the specified f(AGN)$_{\rm MIR}$ refers to the fraction of AGN light to the integrated mid-infrared (i.e., 5-20 $\mu$m) emission. A significant amount of AGN contribution to the MIR SEDs is required by all ultra-massive galaxies at $z>3$ to explain the IR SEDs. If 3$\sigma$ upper limits for the $Herschel$ photometry are adopted, we find a lower limit on the fraction of AGN to the MIR SEDs ranging from $15\%$ to $50\%$ depending on the galaxy. If the 1$\sigma$ upper limits are adopted, we find a lower limit on the fraction of AGN to the MIR SEDs ranging from $40\%$ to $75\%$. 
Columns 5-8 of Table~\ref{tab-ir} list the values or upper limits of L$_{\rm IR}$ and SFR$_{\rm IR}$, and the values or lower limits of $f_{\rm Bol}$(AGN) and $f_{\rm MIR}$(AGN) as obtained from adopting the best-fit templates from \citet{dale14}. The upper limits were derived adopting the 3$\sigma$ upper limits in the {\it Herschel} photometry.
We stress that these newly derived $L_{\rm IR}$ and SFRs should be considered as upper limits, given that we only have lower limits on the AGN contribution to the IR. We note that the three galaxies with X-ray detections, namely A2-15753, C1-15182, and C1-19536 (see Section~\ref{sec-agncont} below), are undetected in $Herschel$ while being detected at several $\sigma$ in the MIPS 24$\mu$m band, f(AGN)$_{\rm MIR}>40\%(75\%)$, $50\%(80\%)$, and $15\%(60\%)$, respectively, when adopting the 3$\sigma$~(1$\sigma$) upper limits in $Herschel$. One target, C1-21316, which has f(AGN)$_{\rm MIR}\sim0.35-0.50$ is detected in radio data, indicative of hosting a radio-loud AGN (Section~\ref{sec-xrad}). 

We also used the templates from the observed FIR color-based library of \citet{kirkpatrick15} to investigate the AGN continuum contribution to the IR SEDs. In this study, libraries of empirical IR SED templates were created from a large sample of high redshift ULIRGS with rest-frame mid-IR spectroscopy from the {\it Spitzer Space Telescope} Infrared Spectrograph (IRS). We display their color diagnostic in the bottom right panel of Figure~\ref{fig-ir}, with the FIR colors of our targets calculated using photometry from {\it Herschel} SPIRE and {\it Spitzer} MIPS/IRAC plotted in solid colored circles (color scheme identical to Figures~\ref{fig-oldnewpar} and \ref{fig-uvj}), with arrows indicating 2$\sigma$ limits. \citet{kirkpatrick15} used mid-IR features to determine the strength of the AGN contribution to IR SEDs by jointly fitting the mid-IR spectrum of a prototypical starburst (M82) and a pure power law for the AGN component to the rest-frame mid-IR spectroscopy. FIR color-based templates were created by stacking the SEDs of galaxies that fall within each $S_{250}/S_{24}$ (observed) vs. $S_{8}/S_{3.6}$ (observed) color box ($<f_{\rm AGN}>$ of galaxies that fall into each color box is indicated in the lower right panel of Figure~\ref{fig-ir}). The \citet{kirkpatrick15} color-based templates consistent with SPIRE, MIPS and IRAC fluxes and detection limits of our targets are displayed in violet in Figure~\ref{fig-ir}, scaled to the MIPS 24$\mu$m flux of each galaxy. Using the appropriate color-based templates and the listed fraction of $L_{\rm IR}$ attributable to star formation for each template (Table 3 in \citealt{kirkpatrick15}), we re-calculated the dust-enshrouded SFRs. The SFRs calculated accounting for the AGN contribution to the FIR SEDs this way are broadly consistent with the observed UV-8$\mu$m SED derived SFRs. Specifically, we calculated the dust-enshrouded SFRs for C1-15182, C1-19536 and A2-15753 to be $<100$ M$_{\odot}$yr$^{-1}$, $\sim 100-200$ M$_{\odot}$yr$^{-1}$ for C1-21316, and $\sim 200$ M$_{\odot}$yr$^{-1}$ for C1-2127, in line with SED derived SFRs within uncertainties. It should be noted that the \citet{kirkpatrick15} library is more representative of our targets as the templates were created using a large sample of ULIRGs at higher redshifts ($z\sim 0.3 - 2.8$) than compared to \citet{dale14} templates, which used local galaxies. 

To summarize, the investigation of the FIR SEDs of the massive galaxy sample reveals that none of them can be adequately described as purely star-forming systems. In fact, a MIR AGN contribution of at least $\sim 40\%$, on average, is necessary to explain the marginal detections in the {\it Herschel} bands. It is promising that estimates of AGN contribution using the template sets of \citet{dale14} and \citet{kirkpatrick15} are consistent with each other. Although in both libraries the fractional AGN contribution is calculated over the rest-frame MIR spectrum, the quasar and star-burst templates differ. \citet{dale14} uses a single quasar template (median PG quasar spectrum of \citealt{shi13}) to linearly mix with a suite of local normal star-forming galaxies spanning a range in $\alpha$ to characterize different heating levels ($\alpha=1$ and 2.5 for active and quiescent galaxies, respectively) over the 5-20$\mu$m wavelength range. In \citet{kirkpatrick15}, the mid-IR spectrum of the prototypical starburst M 82 is used to represent the star-formation component, and the AGN component is characterized by a pure (free-slope) power-law.

\subsubsection{[OIII] Luminosities and Bolometric Correction}
We use the luminosity-dependent O{\small[III]} bolometric correction factor, $C_{\rm OIII}$ from \citet{lamastra09} for the two galaxies with [OIII] emission line detections (namely C1-19536 and C1-15182), $L_{\rm bol}\approx 454\times L_{\rm OIII}$. As both C1-19536 and C1-15182 have non-negligible best-fit $A_{\rm V}$ values from SED fitting, we assume that the observed [OIII] line luminosities are a lower limit to the intrinsic luminosity of [OIII]. Assuming all the observed [OIII] emission is due to AGN radiation, we find $L_{\rm bol}>8\times10^{45}$ erg/s and $>3.5\times10^{45}$ erg/s for C1-19536 and C1-15182, respectively, consistent with them hosting powerful hidden AGNs. 

\subsubsection{Continuum emission from AGN}\label{sec-agncont}

\begin{deluxetable}{lcccc}
\centering
\tablecaption{Best-fit stellar mass and AGN contributions with {\textsc AGNfitter} \label{tab-agnfit}}
\tablehead{ \colhead{ID} & \colhead{$\log M_*$} & \colhead{$\Delta \log M_{*}$} & \colhead{$f_{\rm AGN}$(0.1-1$\mu$m)} & \colhead{$f_{\rm AGN}$(1-30$\mu$m)} } 
\startdata
C1-2127 & $11.24^{+0.17}_{-0.24}$	     & $-0.11^{+0.17}_{-0.25}$	& $0.37^{+0.07}_{-0.07}$ 		&  $0.09^{+0.18}_{-0.05}$  \\
C1-15182 &  $11.61^{+0.05}_{-0.09}$    &	$ +0.05^{+0.17}_{-0.13}$	&  $0.09^{+0.12}_{-0.07}$		& $0.96^{+0.04}_{-0.17}$\\
C1-19536 &  $11.37^{+0.03}_{-0.04}$    &	$-0.08^{+0.10}_{-0.10}$	&  $0.10^{+0.10}_{-0.07}$ 	& $0.98^{+0.02}_{-0.21}$\\
C1-21316 &  $ 11.34^{+0.23}_{-0.68} $  & $-0.24^{+0.57}_{-0.68}$	&  $0.28^{+0.35}_{-0.23}$		& $0.39^{+0.04}_{-0.05}$ \\
A2-15753 &  $11.21^{+0.01}_{-0.03}$    & $+0.04^{+0.15}_{-0.13}$	&  $0.04^{+0.03}_{-0.02}$ 	& $0.99^{+0.01}_{-0.01}$ \\

\enddata
\tablecomments{Results obtained from the modeling of the binned UV-NIR spectra, NMBS UV-8$\mu$m photometry and FIR detections and upper limits with \textsc{AGNfitter}. The best-fit stellar masses were adjusted to the \citet{kroupa01} IMF in order to compare with results listed in Table~\ref{tab-sed}. $\Delta M_{*}$ lists the effect on the derived stellar mass when including AGN templates in the modeling of the panchromatic SEDs ($\Delta \log M_{*}= \log(M_{*})_{({\rm AGNfitter})}- \log(M_{*})_ {({\rm FAST})}$) at fixed metallicity ($Z=Z_{\odot}$). $f_{AGN}$ is calculated as the relative contribution of the AGN templates in the considered wavelengths (i.e, the Big Blue Bump emission describing the accretion disk at $0.1-1\mu$m and the hot dusty torus emission at $1-30\mu$m). }
\end{deluxetable}

\begin{figure*}
\includegraphics[width=\textwidth]{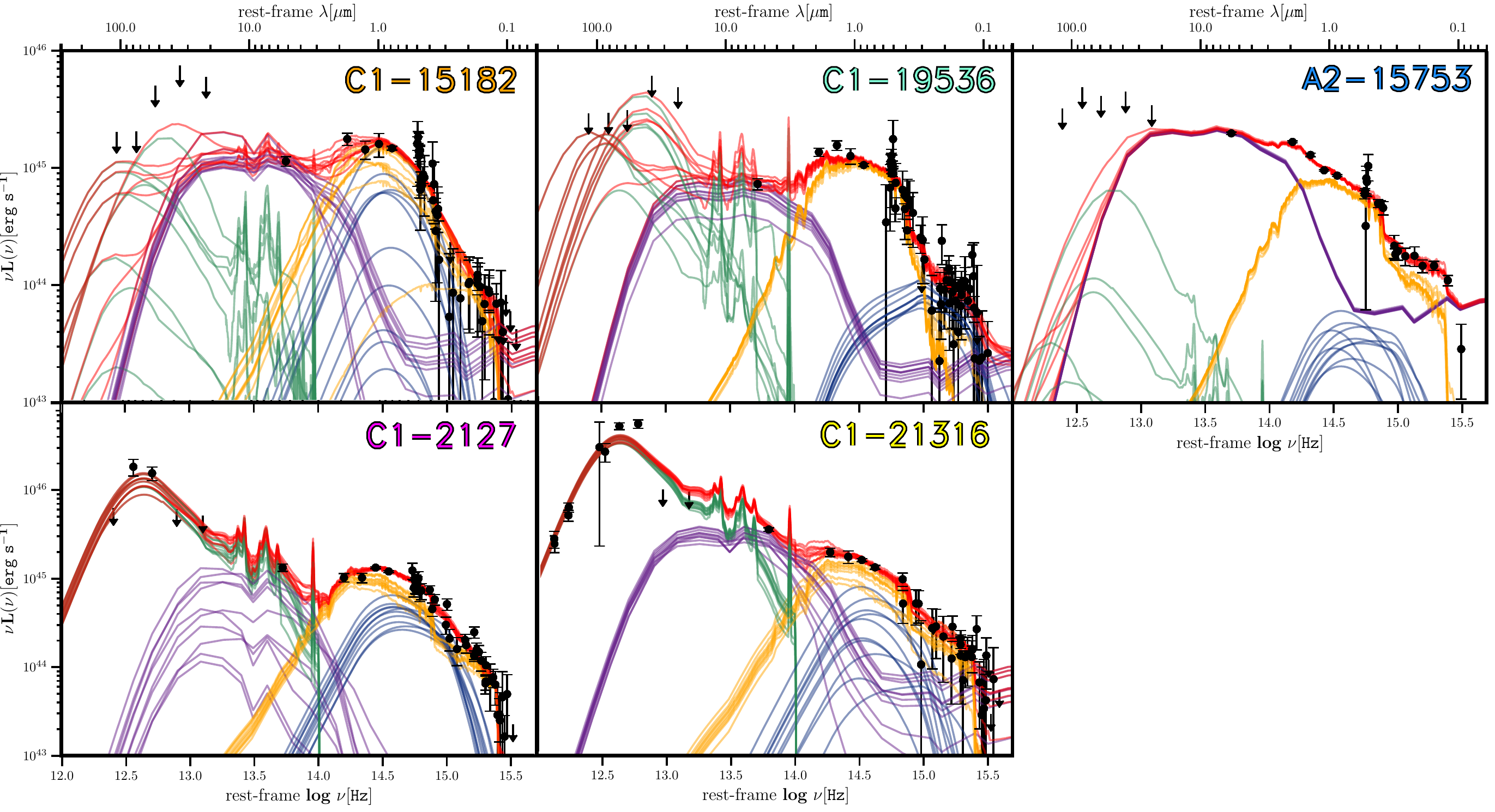}
\caption{The SEDs output when modeling the observed UV-FIR photometry of the 24$\mu$m detected targets with four components using \textsc{AGNfitter} \citep{calistro16}. The orange and green curves correspond to the stellar emission and starburst components, respectively. The accretion disk and the surrounding hot dusty torus components are represented as blue and purple curves, respectively. In each panel, the SED components corresponding to ten randomly selected realizations from the posterior probability density functions are over-plotted in order to visualize the range of parameter space. The total SEDs from the realizations are plotted in red.  \label{fig-agnSED}}
\end{figure*}

The stellar population parameters derived in Section~\ref{sec-sedmodel} may be influenced by the presence of strong AGN continuum contamination to the observed photometry, most significantly biasing the stellar mass estimates. 
We investigated this by using the publicly available fully Bayesian Markov Chain Monte Carlo SED fitting algorithm of active galaxies, {\textsc AGNfitter}
 \footnote{\url{https://github.com/GabrielaCR/AGNfitter}} \citep{calistro16}, to model the observed SEDs of the 24$\mu$m detected targets. 
 {\textsc AGNfitter} simultaneously fits the total active galaxy SED by decomposing it into four physically motivated components: 
the stellar population of the host galaxy, cold dust emission from star forming regions, the accretion disk emission
 (Big Blue Bump, BBB) and hot dust emission surrounding the accretion disk (torus). We used the fiducial library of templates that are supplied with \textsc{AGNfitter}, which has been tested 
 to perform well in classifying Type1 and Type 2 AGN purely based on observed broadband photometry. 
 We briefly describe the modeling assumptions (as adopted in \citealt{calistro16}) and the fiducial library of templates used in \textsc{AGNfitter}, and highlight important distinctions compared to the modeling assumptions employed in previous sections.

The BBB describing the thermal radiation emitted from the accretion disk surrounding the central supermassive black hole was modeled using a single modified template based on the composite spectrum of Sloan Digial Sky Survey Type 1 QSOs \citep{richards06}. Extinction to the emitted BBB spectrum was modeled assuming a Small Magellanic Cloud reddening law \citep{prevot84}. Emission from the dusty torus was modeled by using a library of SEDs with varying hydrogen column densities ($N_{H}=21-25$) created based on the set of empirical templates of \citet{silva04}. The BC03 stellar population synthesis models are used to construct the library of templates to describe the host galaxy's stellar emission. Although the stellar population synthesis model assumptions are similar to those adopted in FAST as presented in Section~\ref{sec-sedmodel}, the explored parameters space is somewhat reduced. Specifically, a similar exponential declining star-formation history is adopted in \textsc{AGNfitter}, of the form $SFR(t)\propto e^{-t/\tau}$, with a mildly more restricted range for the timescale $\tau$ from 0.1 to 10 Gyr. A model with constant SFR is also included in \textsc{AGNfitter}. The grid of stellar population ages is created from 0.2 Gyr to the age of the universe at the target's redshift in step of $\Delta t \sim 0.1$ dex at fixed solar metallicity. \textsc{AGNfitter} adopts a \citet{chabrier} and the \citet{calzetti00} reddening law to model the extinction to the stellar light.
The library of templates for modeling the emission from cold dust in star-forming regions includes the template set of \citet{dale02} (also used in Section~\ref{sec-irsed}) and \citet{chary01}. For a more detailed description of the fitting algorithm, modeling assumptions and template libraries, we refer the reader to \citet{calistro16}.

Figure~\ref{fig-agnSED} illustrates the results when modeling the observed SEDs with the full parameter space range of the fiducial templates of \textsc{AGNfitter} and Table~\ref{tab-agnfit} lists the best-fit stellar masses converted to the \citet{kroupa01}. The fractions of AGN emission in the rest-frame UV-optical ($0.1-1\mu$m) and MIR ($1-30\mu$m) for each object, calculated by comparing the relative contribution of AGN luminosities in the considered wavelenghts are also listed in Table~\ref{tab-agnfit}.
The amount of AGN contribution to the rest-frame MIR estimated by \textsc{AGNfitter} is remarkably consistent with the values or lower limits derived in Section~\ref{sec-irsed}. The AGN contribution in the rest-frame UV-optical is found to range between a few percent (for A2-15753) to $\sim40\%$ (for C1-2127), whereas the AGN contribution in the rest-frame MIR is found to range between $\sim10\%$ (for C1-2127) to almost $100\%$ (for C1-15182, C1-19536, and C1-21316).

The third column of Table~\ref{tab-agnfit} lists the difference between the stellar mass estimated by \textsc{AGNfitter} and the stellar mass derived using FAST (i.e., assuming no AGN contamination to the observed SEDs). It is interesting to note that, while the naive expectation would be that not accounting for AGN contamination to the observed photometry would lead to over-estimating the stellar masses, two galaxies (C1-15182 and A2-15753) have \textsc{AGNfitter} derived stellar masses marginally larger (by $\sim0.04-0.05$ dex) than when no AGN contamination is assumed. This appears to be caused by slightly older best-fit ages of the stellar population (and hence slightly larger mass-to-light ratios) when the SEDs are modeled accounting for the AGN contribution. We note however that these differences are smaller than (or comparable to) the 1$\sigma$ error on the stellar mass. For the other three galaxies, the stellar masses derived by accounting for the AGN contamination are found smaller (by $\sim0.1-0.3$ dex on average) than the stellar masses derived when no AGN contamination is assumed. The best-fit stellar masses derived assuming AGN contamination ranges from  $1.5\times10^{11}~M_{\odot}$ to $4\times10^{11}~M_{\odot}$. All galaxies except C1-21316 are found to be more massive than 10$^{11}$~M$_{\odot}$ within 1$\sigma$ when modeled by \textsc{AGNfitter}. For all galaxies but C1-21316, the systematic effect to the derived stellar masses from AGN contamination is found to be smaller than a factor of $\sim$2 at 1$\sigma$. This is because the rest-frame UV-optical SEDs of these galaxies are dominated by the stellar light despite the rest-frame MIR SEDs being dominated by the obscured AGN. For C1-21316, the stellar mass is found to be overestimated by  a factor of $\sim$2 on average, although the 1$\sigma$ uncertainty allows for solutions smaller by as much as a factor of $\sim$8. Finally, we note that the stellar masses of all studies galaxies derived from \textsc{AGNfitter} are consistent with the stellar masses derived using FAST within the 1$\sigma$ uncertainties. 

\subsubsection{X-ray and Radio Detections} \label{sec-xrad}
As previously done in \citet{marchesini10}, we used the publicly available Chandra X-ray catalogs (\citealt{laird09} and \citealt{elvis09}
for the AEGIS and COSMOS fields, respectively) and the improved redshifts to revise the X-ray luminosities of the three detected sources
(C1-15182, C1-19536, and A2-15753). Assuming a power-law photon index $\Gamma$ = 1.9 \citep{nandrapounds94}, the X-ray luminosities $L_{2-10 keV}$
are (6.4 $\pm$ 1.7)$\times$10$^{45}$ for C1-15182, (16.6 $\pm$ 2.6)$\times$ 10$^{45}$ for C1-19536 and (9.6 $\pm$ 1.5)$\times$ 10$^{45}$ erg~s$^{-1}$
for A2-15753 (redshift uncertainties are included in quoted errors). We used the empirically derived observed $L_{\rm[OIII]}-L_{\rm 2-10 keV}$ relation of local AGNs from \citet{ueda15} to investigate the expected X-ray flux of the sources with detected [OIII] emission (C1-19536 and C1-15182). The estimated $L_{\rm 2-10 keV}$ for these galaxies using the observed $L_{\rm [OIII]}$ are an order of magnitude less than the luminosities calculated using X-ray detections (using $\Gamma=1.4$ lowers this factor to $\sim6-8$). We calculated upper limits to the rest-frame $L_{\rm 2-10 keV}$  for the targets not detected (C1-2127, C1-19764 and C1-21316) using the 3$\sigma$ detection limits with $\Gamma$ = 1.9. The 3$\sigma$ upper limits are in the range $L_{\rm 2-10 keV} \sim 5-20\;\times10^{44}$ erg~s$^{-1}$, comparable to only the very powerful X-ray AGN, and therefore we cannot conclude or rule out whether these galaxies harbor AGN.

From the sample of galaxies presented here, only C1-21316 (AzTEC-5) is detected at 1.4GHz \citep{schinnerer10}, with a flux of 0.126$\pm$0.015 mJy. We calculated the rest-frame flux densities assuming the canonical value of the radio spectral index of $\alpha=0.8$ \citep{condon92}. The calculated rest-frame flux densities is $L_{\rm 1.4GHz}$= 10.6$\pm$1.3 $\times 10^{27}$ W/Hz, above the threshold to be classified as radio-loud AGN (log($L_{\rm 1.4GHz}$>25), \citealt{schinnerer07}). The radio spectral index is not observed to  evolve significantly over cosmic time \citep{magnelli15}, and is consistent with values derived for different samples of galaxies \citep{ibar09, ibar10, ivison10, bourne11}. 

\subsubsection{Redshift Evolution of AGN Fraction}\label{sec-zfagn}

\begin{figure}
\epsscale{1.2}
\plotone{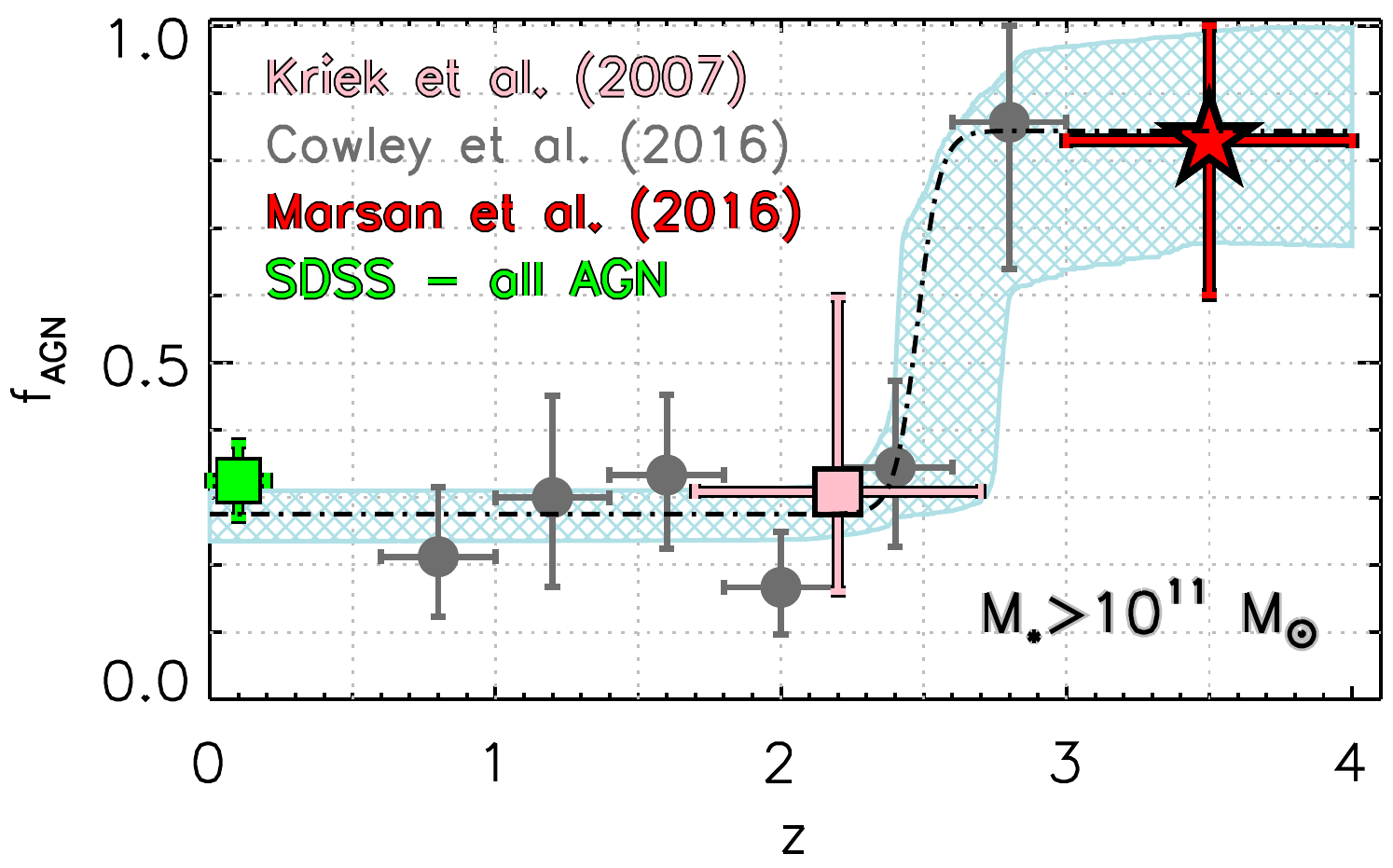}
\caption{Evolution of the fraction of AGNs in galaxies with stellar mass $M_{\star}>10^{11}$~M$_{\odot}$. The red star represents the measurement from our work. The green, pink, and gray points represent the fraction of AGNs at $z\sim0$ (SDSS; \citealt{kauffmann03}), at $1.7<z<2.7$ from \citet{kriek07}, and at $0.6<z<3.0$ from \citet{cowley16}. The dot-dashed curve represents a fit to all plotted points using the empirical model $f_{\rm AGN}=\frac{1}{C+e^{-(z-A)/B}} + D$ with the 1$\sigma$ uncertainties on the model fit indicated by the light blue hatched region. \label{fig-fagn}}

\end{figure}

The fraction of AGN ($f_{\rm AGN}$) within a galaxy population represents a probe of the level of AGN activity and of the growth of super-massive black holes at the center of galaxies. The energetic output from AGN is considered one of the dominant feedback processes that can regulate, and even halt altogether, the infalling of gas and star formation in massive galaxies. The measurement of the level of AGN activity in massive galaxies as a function of redshift is therefore an indirect probe of the impact of AGN feedback at a given cosmic time. 

Figure~\ref{fig-fagn} shows the evolution of $f_{\rm AGN}$ as a function of redshift for galaxies with stellar mass $M_{\star}>10^{11}$~M$_{\odot}$. In our sample of very massive galaxies at $3<z<4$, AGNs appear almost ubiquitous, with a very large AGN fraction of $f_{\rm AGN}\approx0.8\pm0.2$ (in which we assumed that C1-2127 does not host an AGN). Figure~\ref{fig-fagn} also shows the fraction of AGN in massive galaxies at $0.6<z<3.0$ from \citet{cowley16}, at $1.7<z<2.7$ from \citet{kriek07}, and in the local universe from \citet{kauffmann03}. 
Our measurement is consistent with the very high AGN fraction of $f_{\rm AGN}=0.86^{+0.14}_{-0.22}$ at $2.6<z<3.0$ as measured by \citet{cowley16} using the zFOURGE dataset. On the contrary, the AGN fraction in massive galaxies at $z\sim0$ is much smaller, $f_{\rm AGN}\approx0.3$, and it appears to remain relatively constant all the way out to $z\sim2.5$. 
Figure~\ref{fig-fagn} clearly shows a dramatic transition in the fraction of AGNs in galaxies with $M_{\star}>10^{11}$~M$_{\odot}$ happening at $z\sim2.5$, i.e., when the universe was $\sim$3~Gyr old. 
We quantified the significance of the transition in the AGN fraction at $z\sim2.5$ by assuming that the fraction of AGN is constant with redshift, and equal to the bi-weight mean of the observed fraction at $z<2.5$ (i.e., $f_{\rm AGN}\approx0.29$) and testing this hypothesis with the chi-squared statistics of the two points at $z>2.5$. If each point at $z>2.5$ is considered by itself, we find a probability $p<0.0006$ of obtaining each $z>2.5$ measurement at least this discrepant from the no evolution assumption. If the two $z>2.5$ measurements of the AGN fraction are considered together, we find a probability $p<10^{-7}$ of obtaining both $z>2.5$ points at least this discrepant from the no evolution assumption. 
At earlier times, most (all) supermassive black holes at the center of massive galaxies appear to be actively accreting, whereas at later times most ($\sim$2/3) of them are found dormant. 
Although the small sample does not allow us to robustly quantify a potential enhancement of the AGN fraction in our spectroscopically confirmed sample of very massive galaxies at $z>3$ due to [OIII] emission line contamination to the stellar continuum. However, we note that the [OIII] line emissions detected in the spectra are found to contribute at most $15\%$, and more generally $\leq 5\%$, to the observed K-band fluxes. The limited observed [OIII] emission line contamination suggests that our sample is not significantly biased toward strong [OIII] emitters, i.e., toward massive galaxies hosting AGNs, and that the derived AGN fraction in our sample is not considerably enhanced.
Future wide area surveys will be able to constrain more robustly the fraction of AGN in massive galaxies at $z>2.5$, whereas deeper surveys will be able to investigate the evolution of the AGN fraction in lower-mass galaxies.


\section{Summary and Discussion} \label{sec-disc}
In this paper we investigated the observed-frame UV and NIR spectra of a sample of six $3 \leq z_{phot}<4$ massive galaxies selected from the NMBS. We confirmed the spectroscopic redshift of two galaxies (C1-15182, $z_{\rm spec}=3.371$; C1-19536, $z_{\rm spec}=3.139$) through nebular emission lines (Figure~\ref{fig-lines}). We re-modeled the medium- and broadband photometry in combination with the  binned one-dimensional spectra of detected sources to derive improved photometric redshifts with the EAZY code ($z_{\rm cont}$). The best-fit $z_{\rm cont}$ of only one of our targets (C1-19764, not included in \citealt{marchesini10} sample to calculate the high-mass end SMF at $3\leq z_{phot}<4$) is $z<3$ (although the redshift solution extends to $z>3$). 

We used FAST in conjunction with the BC03 stellar population synthesis models, the \citet{calzetti00} extinction law, the \citet{kroupa01} IMF and an exponentially declining SFH to model the observed SEDs (Figure~\ref{fig-fastsed}, Table~\ref{tab-sed}). We included an additional source in our analysis from the \citet{marchesini10} sample, C1-21316, which has a spectroscopic redshift of $z_{\rm spec}=3.971$. We find that the updated stellar population parameters are consistent with those previously derived using only medium- and broadband photometry. The stellar masses of galaxies with detected optical emission lines decrease by $\sim0.1$ dex when emission line contamination to the observed photometry is accounted for. From the SED modeling, our sample of massive galaxies show ranges in stellar population properties, in accordance with previous studies at $3\leq z<4$ \citep{muzzin13b, marchesini14, spitler14}. We find that $\sim50\%$ of sources in this sample are characterized by average stellar ages of $\approx 800$~Myr, suggesting that the bulk of the stellar mass in these systems was formed early in the universe ($z_{\rm form}\sim 4.5$) (these also have ongoing star formation, with SFR$\sim$few$\times$10 - 100 M$_{\odot}$yr$^{-1}$, but these values are highly dependent on model parameters).

All but one of the sources in this sample have significant MIPS 24$\mu$m detections. The total IR luminosities estimated from the observed 24 $\mu$m assuming IR templates representative of starburst galaxies are $\sim$2-3 orders of magnitude greater than the SFRs estimated from SED modeling. Inspecting the observed FIR SEDs including $Herschel$-PACS and SPIRE detections and upper limits effectively rules out dust-enshrouded star formation as the only source of the observed 24$\mu$m flux. In fact, using two separate IR libraries that include different levels of contribution from an AGN \citep{dale14,kirkpatrick15} reveals that $>15\%$ of the MIR emission must originate from obscured AGN. Yet, despite the significant AGN emission contribution to the MIR SEDs, we find that the rest-frame UV-optical light is largely dominated by stellar emission. For all but one galaxy, the presence of the AGN does not considerably bias the stellar mass estimates, with typical systematic differences within $\sim$0.1~dex, and stellar masses at most a factor of $\sim$2 smaller at the 1$\sigma$ level. Even after accounting for the presence of the (mostly obscured) AGN, the inferred stellar masses of all one but galaxy are found to be larger than 10$^{11}$~M$_{\odot}$ within the 1$\sigma$ uncertainties.
The large fraction ($>80\%$) of massive galaxies hosting AGN at $3<z<4$ inferred from strong [OIII] emission lines, IR SED properties, X-ray and radio detections implies that $3<z<4$ must be an extremely active time in the cosmic history for the growth of the super-massive black holes at the massive end of the galaxy population. 

\acknowledgments
The authors thank the anonymous referee for the useful comments and suggestions, which have improved this manuscript. 
ZCM gratefully acknowledges support from the John F. Burlingame and the 
Kathryn McCarthy Graduate Fellowships in Physics at Tufts University. DM and ZCM 
acknowledge the support of the Research Corporation for Science Advancement's 
Cottrell Scholarship, the support from Tufts University Mellon Research 
Fellowship in Arts and Sciences. We thank Gabriela Calistro Rivera for her support on the implementation of \textsc{AGNfitter}. 
This study is partially based on data products from observations made with ESO Telescopes at the La Silla Paranal Observatory under programme ID 087.A-0514. Some of the data presented in this study were obtained at the W.~M.~Keck Observatory, which is operated as a scientific partnership among the California Institute of Technology, the University of California, and the National Aeronautics and Space Administration. The Observatory was made possible by the generous financial support of the W.~M.~Keck Foundation. Keck telescope time was granted by NOAO, through the Telescope System Instrumentation Program (TSIP). TSIP is funded by NSF. The authors wish to recognize and acknowledge the very significant cultural role and reverence that the summit of Mauna Kea has always had within the indigenous Hawaiian community.  We are most fortunate to have the opportunity to conduct observations from this mountain. This study makes use of data from the NEWFIRM Medium-Band Survey, a multi-wavelength survey conducted with the NEWFIRM instrument at the KPNO, supported in part by the NSF and NASA. We thank the NMBS and COSMOS collaborations for making their catalogs publicly available. The combination of NEWFIRM Medium-Band Survey and Herschel data was supported by the National Aeronautics and Space Administration (NASA) under grant number NNX13AH38G issued through the NNH12ZDA001N Astrophysics Data Analysis Program (ADAP).  

\end{document}